\documentclass[12pt, draftclsnofoot, onecolumn]{IEEEtran}
\usepackage{graphicx,subfigure,psfrag,amsmath,cases}
\usepackage{latexsym,amsmath,amssymb,subfigure,algorithm,mathtools}
\usepackage{algorithmic}
\usepackage{color}
\usepackage{url}
\usepackage{scrtime}
\usepackage{cite}
\usepackage{subfigure}
\usepackage{bbding}
\usepackage{multicol}

\author{{Zhiqiang Wei, Lou Zhao, Jiajia Guo, Derrick Wing Kwan Ng, and Jinhong Yuan}
\thanks{Zhiqiang Wei, Lou Zhao, Jiajia Guo, Derrick Wing Kwan Ng, and Jinhong Yuan are with the School of Electrical
Engineering and Telecommunications, the University of New South Wales, Australia (email: zhiqiang.wei@student.unsw.edu.au; lou.zhao@student.unsw.edu.au; jiajia.guo@student.unsw.edu.au; w.k.ng@unsw.edu.au; j.yuan@unsw.edu.au). The conference version of this paper has been presented at the IEEE ICC 2018 \cite{Wei2018mmWaveNOMA}.}}

\title{Multi-Beam NOMA for Hybrid mmWave Systems}

\newtheorem{Def}{Definition}

\newtheorem{T-Prob}{Transformed Problem}

\DeclareMathOperator{\maxo}{maximize}
\DeclareMathOperator{\mino}{minimize}

\newtheorem{Remark}{Remark}

\newcommand{\abs}[1]{\lvert#1\rvert}

\begin{document}
\maketitle
\vspace{-17mm}
\begin{abstract}
In this paper, we propose a multi-beam non-orthogonal multiple access (NOMA) scheme for hybrid millimeter wave (mmWave) systems and study its resource allocation.
A beam splitting technique is designed to generate multiple analog beams to serve multiple users for NOMA transmission.
Compared to conventional mmWave orthogonal multiple access (mmWave-OMA) schemes, the proposed scheme can serve more than one user on each radio frequency (RF) chain.
Besides, in contrast to the recently proposed single-beam mmWave-NOMA scheme which can only serve multiple NOMA users within the same beam, the proposed scheme can perform NOMA transmission for the users with an arbitrary angle-of-departure (AOD) distribution.
This provides a higher flexibility for applying NOMA in mmWave communications and thus can efficiently exploit the potential multi-user diversity.
Then, we design a suboptimal two-stage resource allocation for maximizing the system sum-rate.
In the first stage, assuming that only analog beamforming is available, a user grouping and antenna allocation algorithm is proposed to maximize the conditional system sum-rate based on the coalition formation game theory.
In the second stage, with the zero-forcing (ZF) digital precoder, a suboptimal solution is devised to solve a non-convex power allocation optimization problem for the maximization of the system sum-rate which takes into account the quality of service (QoS) constraint.
Simulation results show that our designed resource allocation can achieve a close-to-optimal performance in each stage.
In addition, we demonstrate that the proposed multi-beam mmWave-NOMA scheme offers a higher spectral efficiency than that of the single-beam mmWave-NOMA and the mmWave-OMA schemes.
\end{abstract}

\section{Introduction}
Recently, the requirements of ultra-high data rate and massive connectivity\cite{wong2017key} have triggered explosive demands of traffic, which has imposed unprecedentedly challenges for the development of the fifth-generation (5G) wireless communications.
In particular, spectrum congestion under 6 GHz in current cellular systems creates a fundamental bottleneck for capacity improvement and sustainable system evolution.
Subsequently, it is necessary and desirable to extend the use of spectrum to high frequency bands, where a wider frequency bandwidth is available, such as millimeter wave (mmWave) bands \cite{Rappaport2013} ranging from 30 GHz to 300 GHz.
On the other hand, multiple access technology is fundamentally important to support multi-user communications in wireless cellular networks.
Although communication systems utilizing microwave bands, i.e., sub-6 GHz, have been widely investigated, the potential multiple access scheme for mmWave communication systems is still unclear.
Meanwhile, non-orthogonal multiple access (NOMA) has been recognized as a promising multiple access technique for the 5G wireless networks due to its higher spectral efficiency and capability to support massive connectivity\cite{Dai2015,Ding2015b,WeiSurvey2016}.
This treatise aims to explore the interwork between the two important techniques via applying the NOMA concept in mmWave communications.

In the literature, two kinds of architectures have been proposed for mmWave communications, i.e., fully digital architecture and hybrid architecture \cite{XiaoMing2017,zhao2017multiuser,GaoSubarray,lin2016energy}.
Specifically, the fully digital architecture requires a dedicated radio frequency (RF) chain\footnote{A RF chain consists of an analog digital converter/digital analog converter (ADC/DAC), a power amplifier, a mixer, and a local oscillator, etc.\cite{lin2016energy}.} for each antenna.
Hence, the tremendous energy consumption of RF chains, and the dramatically increased signal processing complexity and cost become a major obstacle in applying the fully digital architecture to mmWave systems in practical implementations.
In contrast, hybrid architectures, including fully access\cite{zhao2017multiuser} and subarray structures \cite{GaoSubarray}, provide a feasible and compromise solution for implementing mmWave systems which strike a balance between energy consumption, system complexity, and system performance.
In particular, for the fully access hybrid structures, each RF chain is connected to all antennas through an individual group of phase shifters \cite{zhao2017multiuser}.
For subarray hybrid structures\cite{GaoSubarray}, each RF chain has access to only a disjoint subset of antennas through an exclusive phase shifter for each antenna.
In essence, the two kinds of hybrid structures separate the signal processing into a digital precoder in baseband and an analog beamformer in RF band.
The comparison of the two types of hybrid structures can be found in \cite{lin2016energy}.
In general, the hybrid mmWave architectures are practical for implementation due to the promising system performance, which is also the focus of this paper.
Lately, most of existing works \cite{zhao2017multiuser,GaoSubarray,lin2016energy} have investigated the channel estimation and hybrid precoding design for hybrid mmWave architectures.
However, the design of potential and efficient multiple access schemes for hybrid mmWave systems is rarely discussed.

Conventional orthogonal multiple access (OMA) schemes adopted in previous generations of wireless networks cannot be applied directly to the hybrid mmWave systems, due to the associated special propagation features and hardware constraints\cite{wong2017key,Rappaport2013}.
For instance, in hybrid mmWave systems, an analog beamformer is usually shared by all the frequency components in the whole frequency band.
Subsequently, frequency division multiple access (FDMA) and orthogonal frequency division multiple access (OFDMA) are only applicable to the users covered by the same analog beam.
Unfortunately, the beamwidth of an analog beam in mmWave frequency band is typically narrow with a large antenna array\footnote{The $-3$ dB beamwidth of a uniform linear array with $M$ half wavelength spacing antennas is about $\frac{{102.1}}{M}$ degrees \cite{van2002optimum}.} and hence only limited number of users can be served via the same analog beam.
As a result, the limited beamwidth in practical systems reduces the capability of accommodating multiple users via FDMA and OFDMA, despite the tremendous bandwidth in mmWave frequency band.
Similarly, code division multiple access (CDMA) also suffers from the problem of narrow beamwidth, where only the users located within the same analog beam can be served via CDMA.
Even worse, the CDMA system performance is sensitive to the power control quality due to the near-far effects.
Another OMA scheme, time division multiple access (TDMA), might be a good candidate to facilitate multi-user communication in hybrid mmWave systems, where users share the spectrum via orthogonal time slots.
However, it is well-known that the spectral efficiency of TDMA is inferior to that of non-orthogonal multiple access (NOMA) \cite{Ding2015b,zhang2016energy}.
Moreover, the key challenge of implementing TDMA in hybrid mmWave systems is the requirement of high precision in performing fast timing synchronization since mmWave communications usually provide a high symbol rate.
On the other hand, spatial division multiple access (SDMA) \cite{zhao2017multiuser} is a potential technology for supporting multi-user communications, provided that the base station (BS) is equipped with enough number of RF chains and antennas.
However, in hybrid mmWave systems, the limited number of RF chains restricts the number of users that can be served simultaneously via SDMA, i.e., one RF chain can serve at most one user.
In particular, in overloaded scenarios, i.e., the number of users is larger than the number of RF chains, SDMA fails to accommodate all the users.
More importantly, in order to serve a large number of users via SDMA, more RF chains are required which translates to a higher implementation cost, hardware complexity, and energy consumption.
Thus, the combination of SDMA and mmWave \cite{zhao2017multiuser} is unable to cope with the emerging need of massive connectivity required in the future 5G communication systems\cite{Andrews2014}.
Therefore, this paper attempts to overcome the limitation incurred by the small number of RF chains in hybrid mmWave systems.
To this end, we introduce the concept of NOMA into hybrid mmWave systems, which allows the system to serve more users with a limited number of RF chains.

In contrast to conventional OMA schemes mentioned above, NOMA can serve multiple users via the same degree of freedom (DOF) and achieve a higher spectral efficiency\cite{Ding2014,He2017}.
Several preliminary works considered NOMA schemes for mmWave communications, e.g. \cite{Ding2017RandomBeamforming,Cui2017Optimal, WangBeamSpace2017}.
The authors in \cite{Ding2017RandomBeamforming} firstly investigated the coexistence of NOMA and mmWave via using random beamforming and demonstrated the performance gain of mmWave-NOMA over conventional mmWave-OMA schemes.
Then, based on the random beamforming mmWave-NOMA scheme, the optimal user scheduling and power allocation were designed with the branch-and-bound approach\cite{Cui2017Optimal}.
In \cite{WangBeamSpace2017}, the authors proposed a beamspace multiple-input multiple-output (MIMO) NOMA scheme for mmWave communications by using lens antenna array to improve the spectral and energy efficiency.
However, the proposed schemes in\cite{Ding2017RandomBeamforming,Cui2017Optimal, WangBeamSpace2017} are single-beam mmWave-NOMA schemes where NOMA transmission can only be applied to users within the same analog beam.
In general, when the users within the same analog beam are clustered as a NOMA group and share a RF chain, the single-beam mmWave-NOMA scheme outperforms the conventional mmWave-OMA scheme.
Yet, due to the narrow analog beamwidth in hybrid mmWave systems, the number of users that can be served concurrently by the single-beam mmWave-NOMA scheme is very limited and it depends on the users' angle-of-departure (AOD) distribution.
This reduces the potential performance gain brought by NOMA in hybrid mmWave systems.
On the other hand, very recently, the general idea of multi-beam NOMA, which applies NOMA to multiple users with separated AODs, was proposed and discussed for hybrid mmWave systems in \cite{Xiao2017MultiBeam}.
However, the authors in \cite{Xiao2017MultiBeam} only offered a conceptual discussion while a practical method for realizing multi-beam NOMA is still unknown.
Although the joint power allocation and beamforming design \cite{Zhu2017Joint} was proposed for a pure analog mmWave system, it is not applicable to practical hybrid mmWave systems with multiple RF chains and digital precoders.

In this paper, we propose a multi-beam NOMA framework for a multiple RF chain hybrid mmWave system and study the resource allocation design for the proposed multi-beam mmWave-NOMA scheme.
Specifically, all the users are clustered into several NOMA groups and each NOMA group is associated with a RF chain.
Then, multiple analog beams are formed for each NOMA group to facilitate downlink NOMA transmission by exploiting the channel sparsity and the large scale antenna array at the BS.
To this end, a \emph{beam splitting} technique is proposed, which dynamically divides the whole antenna array associated with a RF chain into multiple subarrays to form multiple beams.
Compared to the conventional single-beam mmWave-NOMA scheme\cite{Ding2017RandomBeamforming,Cui2017Optimal, WangBeamSpace2017}, our proposed multi-beam NOMA scheme offers a higher flexibility in serving multiple users with an arbitrary AOD distribution.
As a result, our proposed scheme can form more NOMA groups, which facilitates the exploitation of multi-user diversity to achieve a higher spectral efficiency.

To improve the performance of our proposed multi-beam mmWave-NOMA scheme, we further propose a two-stage resource allocation design.
In the first stage, given the equal power allocation and applying an identity matrix as the digital precoder, the joint design of user grouping and antenna allocation is formulated as an optimization problem to maximize the conditional system sum-rate.
We recast the formulated problem as a coalition formation game \cite{SaadCoalitionalGame,SaadCoalitional2012,WangCoalitionNOMA,SaadCoalitionOrder,Han2012} and develop an algorithmic solution for user grouping and antenna allocation.
In the second stage, based on the obtained user grouping and antenna allocation strategy, the power allocation problem is formulated to maximize the system sum-rate by taking into account the quality of service (QoS) requirement.
A suboptimal power allocation algorithm is obtained based on the difference of convex (D.C.) programming technique.
Compared to the optimal benchmarks in the two stages, we show that our proposed resource allocation design can achieve a close-to-optimal performance in each stage in terms of the system sum-rate.
Simulation results demonstrate that the proposed multi-beam mmWave-NOMA scheme can achieve a higher spectral efficiency than that of the conventional single-beam mmWave-NOMA scheme and mmWave-OMA scheme.

Notations used in this paper are as follows. Boldface capital and lower case letters are reserved for matrices and vectors, respectively. ${\left( \cdot \right)^{\mathrm{T}}}$ denotes the transpose of a vector or a matrix and ${\left( \cdot \right)^{\mathrm{H}}}$ denotes the Hermitian transpose of a vector or a matrix.
$\emptyset$ denotes an empty set; $\mathbb{C}^{M\times N}$ denotes the set of all $M\times N$ matrices with complex entries; $\mathbb{R}^{M\times N}$ denotes the set of all $M\times N$ matrices with real entries;
$\abs{\cdot}$ denotes the absolute value of a complex scalar or the cardinality of a set; and $\left\| \cdot \right\|$ denotes the $l_2$-norm of a vector.
The circularly symmetric complex Gaussian distribution with mean $\mu$ and variance $\sigma^2$ is denoted by ${\cal CN}(\mu,\sigma^2)$.

\section{System Model}

\begin{figure}[t]
\centering\vspace{-7mm}
\subfigure[The proposed multi-beam mmWave-NOMA scheme.]
{\label{MultiBeamNOMA:a} 
\includegraphics[width=0.65\textwidth]{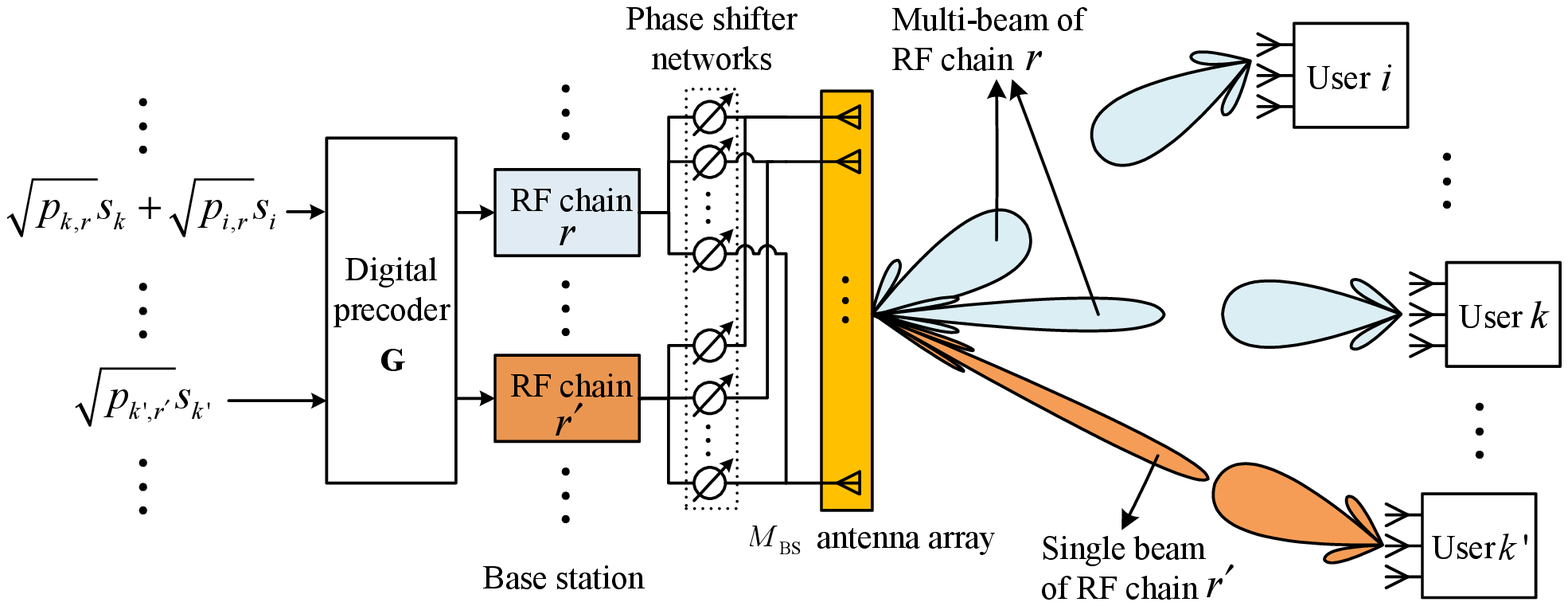}}\vspace{-2mm}
\hspace{1mm}
\subfigure[Hybrid structure receiver at users.]
{\label{MultiBeamNOMA:b} 
\includegraphics[width=0.28\textwidth]{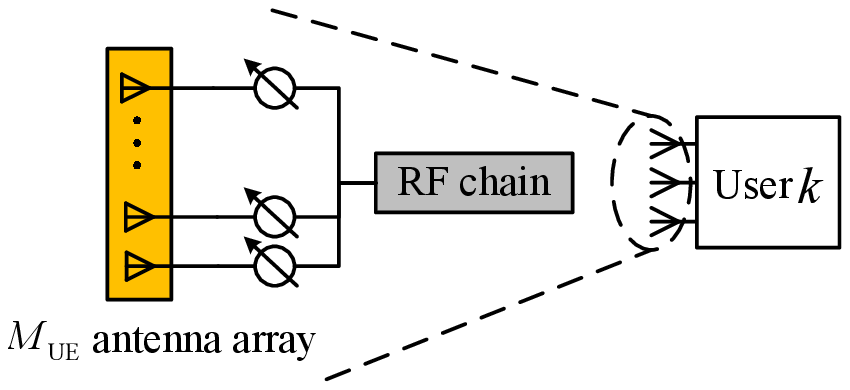}}\vspace{-2mm}
\caption{System model of the proposed multi-beam NOMA scheme for hybrid mmWave systems.}\vspace{-9mm}
\label{MultiBeamNOMA}%
\end{figure}

We consider the downlink hybrid mmWave communication in a single-cell system with one base station (BS) and $K$ users, as shown in Fig. \ref{MultiBeamNOMA}.
In this work, we adopt a fully access hybrid structure\footnote{To simplify the presentation in this paper, the proposed scheme and resource allocation design are based on the fully access hybrid structures as an illustrative example, while they can be easily extended to subarray hybrid structures.} to illustrate the proposed multi-beam NOMA framework for hybrid mmWave systems\cite{zhao2017multiuser,GaoSubarray,lin2016energy}.
In particular, the BS is equipped with $M_{\mathrm{BS}}$ antennas but only connected to $N_{\mathrm{RF}}$ RF chains with $M_{\mathrm{BS}} \gg N_{\mathrm{RF}}$.
We note that each RF chain can access all the $M_{\mathrm{BS}}$ antennas through $M_{\mathrm{BS}}$ phase shifters, as shown in Fig. \ref{MultiBeamNOMA:a}.
Besides, each user is equipped with $M_{\mathrm{UE}}$ antennas connected via a single RF chain, as shown in Fig. \ref{MultiBeamNOMA:b}.
We employ the commonly adopted uniform linear array (ULA) structure\cite{zhao2017multiuser} at both the BS and user terminals.
We assume that the antennas at each transceiver are deployed and separated with equal-space of half wavelength with respect to the neighboring antennas.
In this work, we focus on the overloaded scenario with $K \ge N_{\mathrm{RF}}$, which is fundamentally different from existing works in hybrid mmWave communications, e.g. \cite{zhao2017multiuser,GaoSubarray,lin2016energy}.
In fact, our considered system model is a generalization of that in existing works\cite{zhao2017multiuser,GaoSubarray,lin2016energy}.
For example, the considered system can be degenerated to the conventional hybrid mmWave systems when $K \le N_{\mathrm{RF}}$ and each NOMA group contains only a single user.

We use the widely adopted the Saleh-Valenzuela model \cite{WangBeamSpace2017} as the channel model for our considered mmWave communication systems.
In this model, the downlink channel matrix of user $k$, ${{\bf{H}}_k} \in \mathbb{C}^{{ M_{\mathrm{UE}} \times M_{\mathrm{BS}}}}$, can be represented as
\vspace{-2mm}
\begin{equation}\label{ChannelModel1}
{{\bf{H}}_k} = {\alpha _{k,0}}{{\bf{H}}_{k,0}} + \sum\limits_{l = 1}^L {{\alpha _{k,l}}{{\bf{H}}_{k,l}}},\vspace{-2mm}
\end{equation}
where ${\mathbf{H}}_{k,0} \in \mathbb{C}^{ M_{\mathrm{UE}} \times M_{\mathrm{BS}} }$ is the line-of-sight (LOS) channel matrix between the BS and user $k$ with ${\alpha _{k,0}}$ denoting the LOS complex path gain, ${\mathbf{H}}_{k,l} \in \mathbb{C}^{ M_{\mathrm{UE}} \times M_{\mathrm{BS}} }$ denotes the $l$-th non-line-of-sight (NLOS) path channel matrix between the BS and user $k$ with ${\alpha _{k,l}}$ denoting the corresponding NLOS complex path gains, $1 \le l \le L$, and $L$ denoting the total number of NLOS paths\footnote{If the LOS path is blocked, we treat the strongest NLOS path as ${\mathbf{H}}_{k,0}$ and all the other NLOS paths as ${\mathbf{H}}_{k,l}$.}.
In particular, ${\mathbf{H}}_{k,l}$, $\forall l \in \{0,\ldots,L\}$, is given by
\vspace{-2mm}
\begin{equation}
{\mathbf{H}}_{k,l} = {\mathbf{a}}_{\mathrm{UE}} \left(  \phi _{k,l} \right){\mathbf{a}}_{\mathrm{BS}}^{\mathrm{H}}\left( \theta _{k,l} \right),\vspace{-2mm}
\end{equation}
with ${\mathbf{a}}_{\mathrm{BS}}\left( \theta _{k,l} \right) = \left[ {1, \ldots ,{e^{ - j{\left({M_{{\mathrm{BS}}}} - 1\right)}\pi  \cos \theta _{k,l} }}}\right]^{\mathrm{T}}$
denoting the array response vector \cite{van2002optimum} for the AOD of the $l$-th path ${\theta _{k,l}}$ at the BS and ${\mathbf{a}}_{\mathrm{UE}}\left( \phi _{k,l} \right) = \left[ 1, \ldots ,{e^{ - j{\left({M_{{\mathrm{UE}}}} - 1\right)}\pi \cos \phi _{k,l} }} \right] ^ {\mathrm{T}}$ denoting the array response vector for the angle-of-arrival (AOA) of the $l$-th path ${\phi _{k,l}}$ at user $k$.
Besides, we assume that the LOS channel state information (CSI), including the AODs ${\theta _{k,0}}$ and the complex path gains ${{\alpha _{k,0}}}$ for all the users, is known at the BS owing to the beam tracking techniques \cite{BeamTracking}.
For a similar reason, the AOA ${\phi _{k,0}}$ is assumed to be known at user $k$, $\forall k$.

\section{The Multi-beam NOMA scheme}

\begin{figure}[t]
\centering\vspace{-5mm}
\includegraphics[width=3in]{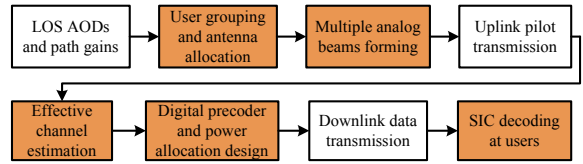}\vspace{-5mm}
\caption{The proposed multi-beam NOMA framework for hybrid mmWave systems. The shaded blocks are the design focuses of this paper.}\vspace{-9mm}
\label{MultiBeamNOMAScheme}
\end{figure}

The block diagram of the proposed multi-beam NOMA framework for the considered hybrid mmWave system is shown in Fig. \ref{MultiBeamNOMAScheme}.
Based on the LOS CSI, we cluster users as multiple NOMA groups and perform antenna allocation among users within a NOMA group.
Then, we control the phase shifters based on the proposed beam splitting technique to generate multiple analog beams.
Effective channel is estimated at the BS based on the uplink pilot transmission and the adopted analog beamformers.
Then, according to the effective channel, the digital precoder and power allocation are designed for downlink data transmission.
Since superposition transmission is utilized within a NOMA group, SIC decoding will be performed at the strong users as commonly adopted in traditional downlink NOMA protocol\cite{WeiSurvey2016}.
The shaded blocks are the design focuses of this paper, which are detailed in the sequel.

\subsection{User Grouping and Antenna Allocation}
Based on the LOS AODs of all the users $\left\{ {\theta _{1,0}}, \ldots ,{\theta _{K,0}} \right\}$ and their path gains $\left\{ {\alpha _{1,0}}, \ldots ,{\alpha _{K,0}} \right\}$, we first perform user grouping and antenna allocation.
In particular, multiple users might be allocated with the same RF chain to form a NOMA group.
We can define the user scheduling variable as follows:
\vspace{-2mm}
\begin{equation}
{u_{k,r}} = \left\{ {\begin{array}{*{20}{c}}
1,&{{\mathrm{user}}\;k\;{\mathrm{is}}\;{\mathrm{allocated}}\;{\mathrm{to}}\;{\mathrm{RF}}\;{\mathrm{chain}}\;r},\\[-1mm]
0,&{{\mathrm{otherwise}}}.
\end{array}} \right.\vspace{-2mm}
\end{equation}
To reduce the computational complexity and time delay of SIC decoding within the NOMA group, we assume that at most $2$ users\footnote{This assumption is commonly adopted in the NOMA literature\cite{WeiTCOM2017,Sun2016Fullduplex}, to facilitate the system resource allocation design.} can be allocated with the same RF chain, i.e., $\sum\nolimits_{k = 1}^K {{u_{k,r}}}  \le 2$, $\forall r$.
In addition, due to the limited number of RF chains in the considered hybrid systems, we assume that each user can be allocated with at most one RF chain, i.e., $\sum\nolimits_{r = 1}^{{N_{{\mathrm{RF}}}}} {{u_{k,r}}}  \le 1$, $\forall k$.
The beam splitting technique proposed in this paper involves antenna allocation within each NOMA group.
Denote $M_{k,r}$ as the number of antennas allocated to user $k$ associated with RF chain $r$, we have $\sum\nolimits_{k = 1}^K u_{k,r}{M_{k,r}}  \le M_{\mathrm{BS}}$, $\forall r$.

\subsection{Multiple Analog Beams with Beam Splitting}
In the conventional single-beam mmWave-NOMA schemes\cite{Ding2017RandomBeamforming,Cui2017Optimal, WangBeamSpace2017}, there is only a single beam for each NOMA group.
However, as mentioned before, the beamwidth is usually very narrow in mmWave frequency bands and a single beam can rarely cover multiple NOMA users, which restricts the potential performance gain brought by NOMA.
Therefore, we aim to generate multiple analog beams for each NOMA group, wherein each beam is steered to a user within the NOMA group.
To this end, we propose the beam splitting technique, which separates adjacent antennas to form multiple subarrays creating an analog beam via each subarray.
For instance, in Fig. \ref{MultiBeamNOMA:a}, user $k$ and user $i$ are scheduled to be served by RF chain $r$ at the BS, where their allocated number of antennas are ${M_{k,r}}$ and ${M_{i,r}}$, respectively, satisfying ${M_{k,r}} + {M_{i,r}} \le {M_{\mathrm{BS}}}$.
Then, the analog beamformer for the ${M_{k,r}}$ antennas subarray is given by
\vspace{-2mm}
\begin{equation}\label{SubarayWeight1}
{\mathbf{w}}\left( {M_{k,r}},{\theta _{k,0}} \right)=
{\frac{1}{{\sqrt {{M_{BS}}} }}{\left[ {1, \ldots ,{e^{j\left( {{M_{k,r}} - 1} \right)\pi \cos  {{\theta _{k,0}}} }}} \right]} ^ {\mathrm{T}}},\vspace{-2mm}
\end{equation}
and the analog beamformer for the ${M_{i,r}}$ antennas subarray is given by
\vspace{-2mm}
\begin{equation}\label{SubarayWeight2}
{\mathbf{w}}\left( {M_{i,r},{\theta _{i,0}}} \right) =
{\frac{{e^{j{M_{k,r}}\pi \cos {{\theta _{k,0}}}}}}{\sqrt{M_{BS}}}{\left[ {1, \ldots ,{e^{j\left( {{M_{i,r}} - 1} \right)\pi \cos {{\theta _{i,0}}} }}} \right]} ^ {\mathrm{T}}},\vspace{-2mm}
\end{equation}
where $j$ is the imaginary unit, ${\mathbf{w}}\left( M_{k,r},{\theta _{k,0}} \right) \in \mathbb{C}^{ M_{k,r} \times 1}$, and ${\mathbf{w}}\left( M_{i,r},\theta _{i,0} \right) \in \mathbb{C}^{ M_{i,r} \times 1}$.
Note that the phase shift term ${e^{j{M_{k,r}}\pi \cos  {{\theta _{k,0}}} }}$ in \eqref{SubarayWeight2} is introduced to synchronize the phase between two subarrays.
Besides, since the phase shift term is imposed on all the elements in the ${M_{i,r}}$ antennas subarray, it does not change the beam pattern response for the ${M_{i,r}}$ antennas subarray.
The same normalized factor $\frac{1}{\sqrt {M_{BS}} }$ is introduced in \eqref{SubarayWeight1} and \eqref{SubarayWeight2} to fulfill the constant modulus constraint \cite{Sohrabi2016} of phase shifters.
As a result, the analog beamformer for RF chain $r$ is given by
\vspace{-2mm}
\begin{equation}\label{ArrayWeight1}
{\mathbf{w}}_r = \left[ {
{{\mathbf{w}}^{\mathrm{T}}\left( {M_{k,r}},{\theta _{k,0}} \right)}, {{{\mathbf{w}}^{\mathrm{T}}}\left( {M_{i,r}},{\theta _{i,0}} \right)}
} \right]^{\mathrm{T}}.\vspace{-2mm}
\end{equation}
On the other hand, if user $k'$ is allocated with RF chain $r'$ exclusively, then all the ${M_{\mathrm{BS}}}$ antennas of RF chain $r'$ will be allocated to user $k'$.
In this situation, the analog beamformer for user $k'$ is identical to the conventional analog beamformer in hybrid mmWave systems, i.e., no beam splitting, and it is given by
\vspace{-2mm}
\begin{equation}\label{ArrayWeight2}
{{\mathbf{w}}_{r'}}= {\mathbf{w}}\left( {{M_{\mathrm{BS}}},{\theta _{k',0}}} \right) =
{\frac{1}{{\sqrt {{M_{BS}}} }}\left[ {1, \ldots ,{e^{j{\left({M_{\mathrm{BS}}} - 1\right)}\pi \cos {{\theta _{k',0}}} }}} \right] ^ {\mathrm{T}}}.\vspace{-2mm}
\end{equation}
Note that, compared to the single-beam mmWave-NOMA schemes\cite{Ding2017RandomBeamforming,Cui2017Optimal, WangBeamSpace2017}, the AODs of the LOS paths ${\theta _{k,0}}$ and ${\theta _{i,0}}$ in the proposed scheme are not required to be in the same analog beam.
In other words, the proposed multi-beam NOMA scheme provides a higher flexibility in user grouping than that of the single-beam NOMA schemes.

\begin{figure}[t]
\centering\vspace{-7mm}
\includegraphics[width=3.5in]{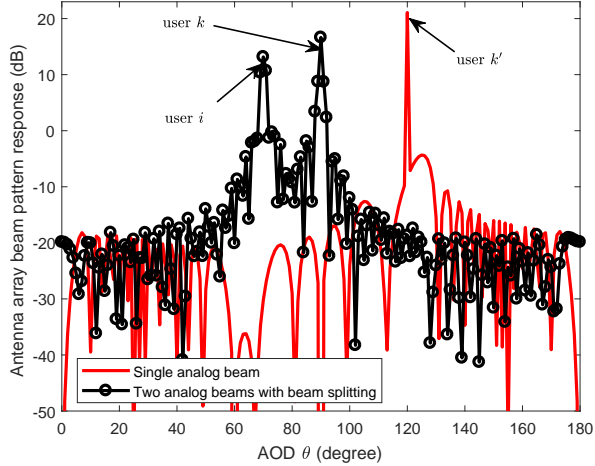}\vspace{-5mm}
\caption{Antenna array beam pattern response for ${{\bf{w}}_r}$ in \eqref{ArrayWeight1} and ${{\bf{w}}_r'}$ in \eqref{ArrayWeight2} at the BS via the proposed beam splitting technique. We assume ${M_{\mathrm{BS}}} = 128$, ${M_{i,r}} = 50$, ${M_{k,r}} = 78$, ${\theta _i} = 70^{\circ}$, ${\theta _k} = 90^{\circ}$, and ${\theta _{k'}} = 120^{\circ}$.}\vspace{-9mm}
\label{MultiBeamNOMA2}
\end{figure}

Based on the analog beamformers ${{\mathbf{w}}_r}$ and ${{\mathbf{w}}_{r'}}$, RF chain $r$ generates two analog beams steering toward user $k$ and user $i$, respectively, while RF chain $r'$  forms a single analog beam steering to user $k'$.
The antenna array beam pattern responses for ${{\mathbf{w}}_r}$ and ${{\mathbf{w}}_{r'}}$ are shown in Fig. \ref{MultiBeamNOMA2} to illustrate the multiple analog beams generated via beam splitting.
Compared to the single analog beam for user $k'$, we can observe that the magnitude of the main beam response decreases and the beamwidth increases for both the two analog beams for users $k$ and $i$.
Although forming multiple analog beams via beam splitting sacrifices some beamforming gain of the original single analog beam, it can still improve the system performance via accommodating more users on each RF chain.

Now, we integrate the users scheduling variables ${u_{k,r}}$ with ${{\mathbf{w}}_r}$ as follows
\vspace{-2mm}
\begin{equation}\label{Weight1}
{{\mathbf{w}}_r} = {\left[
{{{\mathbf{w}}^{\mathrm{T}}}\left( {{u_{1,r}},{M_{1,r}},{\theta _{1,0}}} \right)}, \ldots ,{{{\mathbf{w}}^{\mathrm{T}}}\left( {{u_{K,r}},{M_{K,r}},{\theta _{K,0}}} \right)}
\right]^{\mathrm{T}}},\vspace{-2mm}
\end{equation}
with
\vspace{-2mm}
\begin{equation}\label{Weight2}
{\mathbf{w}}\left( {{u_{k,r}},{M_{k,r}},{\theta _{k,0}}} \right) = \left\{ {\begin{array}{*{20}{c}}
\emptyset, &{{u_{k,r}} = 0},\\[-1mm]
{\frac{{{e^{j\sum\limits_{d = 1}^{k - 1} {{u_{d,r}}{M_{d,r}}\pi \cos \left( {{\theta _{d,0}}} \right)} }}}}{{\sqrt {{M_{BS}}} }}{{\left[ {1, \ldots ,{e^{j\left( {{M_{k,r}} - 1} \right)\pi\cos \left( {{\theta _{k,0}}} \right)}}} \right]}^{\mathrm{T}}}}, &{{u_{k,r}} = 1}.
\end{array}} \right.\vspace{-2mm}
\end{equation}
It can be observed in \eqref{Weight2} that ${\mathbf{w}}\left( {{u_{k,r}},{M_{k,r}},{\theta _{k,0}}} \right)$ is an empty set $\emptyset$ when ${{u_{k,r}} = 0}$, $\forall k$, and ${{\mathbf{w}}_r}$ consists of the analog beamformers for the users allocated with RF chain $r$, i.e., ${{u_{k,r}} = 1}$, $\forall k$.

\subsection{Effective Channel Estimation}
For a given user grouping strategy, antenna allocation, and multiple analog beamforming, all the users transmit their unique orthogonal pilots to the BS in the uplink to perform effective channel estimation.
In this paper, we assume the use of time division duplex (TDD) and exploit the channel reciprocity, i.e., the estimated effective channel in the uplink can be used for digital precoder design in the downlink.
The effective channel of user $k$ on RF chain $r$ at the BS is given by
\vspace{-2mm}
\begin{equation}\label{EffectiveChannel1}
{\widetilde h_{k,r}} = {{\mathbf{v}}_k^{\mathrm{H}}}{{\mathbf{H}}_{k}}{\mathbf{w}}_r,\;\forall k,r,\vspace{-2mm}
\end{equation}
where ${{\mathbf{v}}_k}$ and ${\mathbf{w}}_r$ denote the analog beamformers adopted at user $k$ and the RF chain $r$ at the BS, respectively\footnote{In the proposed multi-beam NOMA framework, any channel estimation scheme can be used to estimate the effective channel in \eqref{EffectiveChannel1}. For illustration, we adopted the strongest LOS based channel estimation scheme \cite{AlkhateebPrecoder2015,zhao2017multiuser} with ${{\mathbf{v}}_k} = \frac{1}{{\sqrt {{M_{{\mathrm{UE}}}}} }}{{\mathbf{a}}_{{\mathrm{UE}}}}\left( {{\phi _{k,0}}} \right)$ and ${\mathbf{w}}_r$ given by \eqref{Weight1}.}.
In the following, we denote the effective channel vector of user $k$ as ${{{\bf{\widetilde h}}}_k} = {\left[ {
{{{\widetilde h}_{k,1}}}, \ldots ,{{{\widetilde h}_{k,{N_{{\mathrm{RF}}}}}}}
} \right]}^{\rm{T}} \in \mathbb{C}^{{ N_{\mathrm{RF}} \times 1}}$ and denote the effective channel matrix between the BS and the $K$ users as ${\bf{\widetilde H}} = {\left[ {
{\widetilde{{\bf{h}}}_1}, \ldots ,{\widetilde{{\bf{h}}}_K}
} \right]} \in \mathbb{C}^{{ N_{\mathrm{RF}}\times K }}$.

\subsection{Digital Precoder and Power Allocation Design}
Given the estimated effective channel matrix ${{\mathbf{\widetilde H}}}$, the digital precoder and the power allocation can be designed accordingly.
With the proposed user grouping design, there are totally $N_{\mathrm{RF}}$ NOMA groups to be served in the proposed multi-beam NOMA scheme and the users within each NOMA group share the same digital precoder.
Assuming that the adopted digital precoder is denoted as ${\mathbf{G}} = \left[ {
{{{\mathbf{g}}_1}},\ldots ,{{{\mathbf{g}}_{{N_{{\mathrm{RF}}}}}}}
} \right] \in \mathbb{C}^{{ N_{\mathrm{RF}} \times N_{\mathrm{RF}}}}$, where ${{\mathbf{g}}_r}$ with ${\left\| {{{\bf{g}}_r}} \right\|^2} = 1$ denotes the digital precoder for the NOMA group associated\footnote{Note that the concept of RF chain association is more clear for a pure analog mmWave system, i.e., ${\mathbf{G}} = {{\mathbf{I}}_{{N_{{\mathrm{RF}}}}}}$, where the signal of a NOMA group or an OMA user is transmitted through its associated RF chain, as shown in Fig. \ref{MultiBeamNOMA:a}.
In hybrid mmWave systems with a digital precoder, ${\mathbf{G}}$, the signals of NOMA groups and OMA users are multiplexed on all the RF chains.
In this case, the RF chain allocation is essentially the spatial DOF allocation, where a NOMA group or an OMA user possesses one spatial DOF.
However, we still name it with RF chain association since each associated RF chain generates multiple analog beams for a NOMA group or a single beam for an OMA user, as shown in Fig. \ref{MultiBeamNOMA:a}.} with RF chain $r$.
In addition, denoting the power allocation for user $k$ associated with RF chain $r$ as $p_{k,r}$, we have the sum power constraint $ \sum_{k = 1}^K \sum_{r = 1}^{{N_{{\mathrm{RF}}}}} {{u_{k,r}}{p_{k,r}}}  \le p_{\mathrm{BS}}$, where $p_{\mathrm{BS}}$ is the power budget for the BS.
Then, the received signal at user $k$ is given by
\vspace{-2mm}
\begin{align}\label{DLRx1}
{y_k} &= {{{\mathbf{\widetilde h}}}_k^{\mathrm{H}}}{{\mathbf{G}}\mathbf{t}} + {z_k} = {{{\mathbf{\widetilde h}}}_k^{\mathrm{H}}}\sum\nolimits_{r = 1}^{{N_{{\mathrm{RF}}}}} {{{\mathbf{g}}_r}{t_r}}  + {z_k} \notag\\[-1mm]
& = \underbrace{{{{\mathbf{\widetilde h}}}_k^{\mathrm{H}}} {{\mathbf{g}}_r} \sqrt {{p_{k,r}}} {s_k}}_{\mathrm{Desired\;signal}} + \underbrace{{{{\mathbf{\widetilde h}}}_k^{\mathrm{H}}} {{\mathbf{g}}_r}\sum\nolimits_{d \ne k}^K {{u_{d,r}}\sqrt {{p_{d,r}}} {s_{d}}}}_{\mathrm{Intra-group\;interference}}
+ \underbrace{{{{\mathbf{\widetilde h}}}_k^{\mathrm{H}}}\sum\nolimits_{r' \ne r}^{{N_{{\mathrm{RF}}}}} {{{\mathbf{g}}_{r'}}\sum\nolimits_{d = 1}^K {{u_{d,r'}}} \sqrt {{p_{d,r'}}} {s_{d}}}}_{\mathrm{Inter-group\;interference}}  + {z_k},
\end{align}
\par
\vspace*{-1mm}
\noindent
where ${t_r} = \sum\nolimits_{k = 1}^K {{u_{k,r}}} \sqrt{p_{k,r}}{s_k}$ denotes the superimposed signal of the NOMA group associated with RF chain $r$ and ${\mathbf{t}} = {\left[
{{t_1}}, \ldots ,{{t_{{N_{{\mathrm{RF}}}}}}}
\right]^{\mathrm{T}}} \in \mathbb{C}^{{ N_{\mathrm{RF}} \times 1}}$.
Variable ${s_k} \in \mathbb{C}$ denotes the modulated symbol for user $k$ and $z_k \sim {\cal CN}(0,\sigma^2)$ is the additive white Gaussian noise (AWGN) at user $k$, where $\sigma^2$ is the noise power.
For instance, in Fig. \ref{MultiBeamNOMA:a}, if user $k$ and user $i$ are allocated to RF chain $r$ and user $k'$ is allocated to RF chain $r'$, we have ${t_r} = \sqrt{p_{k,r}}{s_k} + \sqrt{p_{i,r}}{s_i}$ and ${t_{r'}} = \sqrt{p_{k',r'}}{s_{k'}}$.
In \eqref{DLRx1}, the first term represents the desired signal of user $k$, the second term denotes the intra-group interference caused by the other users within the NOMA group associated with RF chain $r$, and the third term is the inter-group interference originated from all the other RF chains.

\subsection{SIC Decoding at Users}
At the user side, as the traditional downlink NOMA schemes\cite{WeiTCOM2017}, SIC decoding is performed at the strong user within one NOMA group, while the weak user directly decodes the messages by treating the strong user's signal as noise.
In this paper, we define the strong or weak user by the LOS path gain.
Without loss of generality, we assume that the users are indexed in the descending order of LOS path gains, i.e., ${\left| {{\alpha _{1,0}}} \right|^2} \ge {\left| {{\alpha _{2,0}}} \right|^2} \ge , \ldots , \ge {\left| {{\alpha _{K,0}}} \right|^2}$.

According to the downlink NOMA protocol\cite{WeiTCOM2017}, the individual data rate of user $k$ when associated with RF chain $r$ is given by
\vspace{-2mm}
\begin{equation}\label{DLIndividualRate1}
{R_{k,r}} = {\log _2}\left( {1 + \frac{{{u_{k,r}}{p_{k,r}}{{\left| {{{\mathbf{\widetilde h}}}_k^{\mathrm{H}}}{{\mathbf{g}}_r} \right|}^2}}}{{I_{k,r}^{{{\mathrm{ inter}}}} + I_{k,r}^{{{\mathrm{intra}}}} + \sigma^2}}} \right),\vspace{-2mm}
\end{equation}
with $I_{k,r}^{{{\mathrm{inter}}}} = \sum\nolimits_{r' \ne r}^{{N_{{\mathrm{RF}}}}} {{{ \left| { {{{\mathbf{\widetilde h}}}_k^{\mathrm{H}}}{{\mathbf{g}}_{r'}} } \right|}^2}\sum\nolimits_{d = 1}^K {{u_{d,r'}}{p_{d,r'}}} }$ and $I_{k,r}^{{{\mathrm{intra}}}} = {{\left| {{{\mathbf{\widetilde h}}}_k^{\mathrm{H}}}{{\mathbf{g}}_r} \right|}^2}\sum\nolimits_{d = 1}^{k-1} {{u_{d,r}}{p_{d,r}}}$
denoting the inter-group interference power and intra-group interference power, respectively.
Note that with the formulation in \eqref{DLIndividualRate1}, we have ${R_{k,r}} = 0$ if ${u_{k,r}} = 0$.
If user $k$ and user $i$ are scheduled to form a NOMA group associated with RF chain $r$, $\forall i > k$, user $k$ first decodes the messages of user $i$ before decoding its own information and the corresponding achievable data rate is given by
\vspace{-1mm}
\begin{equation}\label{IndividualRate2}
R_{k,i,r} = {\log _2}\left( {1 + \frac{{{u_{i,r}}{p_{i,r}}{{\left| {{{\mathbf{\widetilde h}}}_k^{\mathrm{H}}}{{\mathbf{g}}_r} \right|}^2}}}{{I_{k,r}^{{\mathrm{inter}}} + I_{k,i,r}^{{\mathrm{intra}}} + \sigma^2}}} \right),\vspace{-3mm}
\end{equation}
where $I_{k,i,r}^{{\mathrm{intra}}} = {{\left| {{{\mathbf{\widetilde h}}}_k^{\mathrm{H}}}{{\mathbf{g}}_r} \right|}^2}\sum\nolimits_{d = 1}^{i - 1} {{u_{d,r}}{p_{d,r}}}$ denotes the intra-group interference power when decoding the message of user $i$ at user $k$.
To guarantee the success of SIC decoding, we need to maintain the rate condition as follows \cite{Sun2016Fullduplex}:
\vspace{-2mm}
\begin{equation}\label{DLSICDecoding}
R_{k,i,r} \ge {R_{i,r}}, \forall i > k.\vspace{-2mm}
\end{equation}
Note that, when user $i$ is not allocated with RF chain $r$, we have $R_{k,i,r} = {R_{i,r}} =0$ and the condition in \eqref{DLSICDecoding} is always satisfied.
Now, the individual data rate of user $k$ is defined as $R_{k} = {\sum\nolimits_{r = 1}^{{N_{\mathrm{RF}}}} {{R_{k,r}}}}$, $\forall k$, and the system sum-rate is given by
\vspace{-2mm}
\begin{equation}\label{SumRate}
{R_{{\mathrm{sum}}}} = \sum\limits_{k = 1}^K {\sum\limits_{r = 1}^{{N_{RF}}} {{{\log }_2}\left( {1 + \frac{{{u_{k,r}}{p_{k,r}}{{\left| {{{\mathbf{\widetilde h}}}_k^{\mathrm{H}}}{{\mathbf{g}}_r} \right|}^2}}}{{I_{k,r}^{{{\mathrm{inter}}}} + I_{k,r}^{{{\mathrm{intra}}}} + \sigma^2}}} \right)} }.\vspace{-2mm}
\end{equation}

\begin{Remark}
In summary, the key idea of the proposed multi-beam NOMA framework\footnote{Note that this paper proposes a multi-beam NOMA framework for hybrid mmWave communication systems, where different analog beamformer designs, channel estimation methods, and digital precoder designs can be utilized in the proposed framework.} is to multiplex the signals of multiple users on a single RF chain via beam splitting, which generates multiple analog beams to facilitate non-orthogonal transmission for multiple users.
Intuitively, compared to the natural broadcast in conventional NOMA schemes considered in microwave frequency bands\cite{WeiSurvey2016,Ding2015b}, the proposed multi-beam NOMA scheme generates multiple non-overlapped virtual tunnels in the beam-domain and broadcast within the tunnels for downlink NOMA transmission.
It is worth to note that the beam splitting technique is essentially an allocation method of array beamforming gain.
In particular, allocating more antennas to a user means allocating a higher beamforming gain for this user, and vice versa.
Specifically, apart from the power domain multiplexing in conventional NOMA schemes\cite{WeiSurvey2016,Ding2015b}, the proposed multi-beam NOMA scheme further exploits the beam-domain for efficient multi-user multiplexing.
Besides, the proposed multi-beam NOMA scheme only relies on AODs of the LOS paths, ${\theta _{k,0}}$, and the complex path gains, ${{\alpha _{k,0}}}$, which is different from the existing fully digital MIMO-NOMA schemes\cite{Hanif2016}.
Clearly, the performance of the proposed multi-beam mmWave-NOMA scheme highly depends on the user grouping, antenna allocation, power allocation, and digital precoder design, which will be detailed in the next section.
\end{Remark}

\section{Resource Allocation Design}
In this section, we focus on resource allocation design for the proposed multi-beam mmWave-NOMA scheme.
As shown in Fig. \ref{MultiBeamNOMAScheme}, the effective channel seen by the digital precoder depends on the structure of analog beamformers, which is determined by the user grouping and antenna allocation.
In other words, the acquisition of effective channel is coupled with the user grouping and antenna allocation.
In fact, this is fundamentally different from the resource allocation design of the fully digital MIMO-NOMA schemes\cite{Sun2017MIMONOMA,Hanif2016} and single-beam mmWave-NOMA schemes\cite{Ding2017RandomBeamforming,Cui2017Optimal, WangBeamSpace2017}.
As a result, jointly designing the user grouping, antenna allocation, power allocation, and digital precoder for our proposed scheme is very challenging and generally intractable.
To obtain a computationally efficient design, we propose a two-stage design method, which is commonly adopted for the hybrid precoder design in the literature\cite{AlkhateebPrecoder2015,Mumtaz2016mmwave}.
Specifically, in the first stage, we design the user grouping and antenna allocation based on the coalition formation game theory \cite{SaadCoalitionalGame,SaadCoalitional2012,WangCoalitionNOMA,SaadCoalitionOrder,Han2012} to maximize the conditional system sum-rate assuming that only analog beamforming is available.
In the second stage, based on the obtained user grouping and antenna allocation strategy, we adopt a ZF digital precoder to manage the inter-group interference and formulate the power allocation design as an optimization problem to maximize the system sum-rate while taking into account the QoS constraints.

\subsection{First Stage: User Grouping and Antenna allocation}
\subsubsection{Problem Formulation}
In the first stage, we assume that only analog beamforming is available with ${\mathbf{G}} = {{\mathbf{I}}_{{N_{{\mathrm{RF}}}}}}$ and equal power allocation ${p_{k,r}} = \frac{p_{\mathrm{BS}}}{K}$, $\forall k,r$, i.e., each RF chain serves its associated NOMA group correspondingly.
The joint user grouping and antenna allocation to maximize the achievable sum-rate can be formulated as the following optimization problem:
\vspace{-2mm}
\begin{align} \label{ResourceAllocation}
&\underset{{u_{k,r}},{M_{k,r}}}{\maxo} \;\;\overline{R}_{\mathrm{sum}} = \sum\nolimits_{k = 1}^K \sum\nolimits_{r = 1}^{{N_{RF}}} {\overline{R}_{k,r}} \\[-2mm]
\mbox{s.t.}\;\;
&\mbox{C1: } {u_{k,r}} \in \left\{ {0,1} \right\}, {M_{k,r}} \in \mathbb{Z}{^ + }, \forall k,r,  \;\;\mbox{C2: } \sum\nolimits_{k = 1}^K {{u_{k,r}}}  \le 2, \forall r, \notag\\[-1mm]
&\mbox{C3: } \sum\nolimits_{r = 1}^{{N_{{\mathrm{RF}}}}} {{u_{k,r}}}  \le 1, \forall k, \;\;\mbox{C4: } \sum\nolimits_{k = 1}^K {{u_{k,r}}{M_{k,r}}}  \le {M_{{\mathrm{BS}}}}, \forall r, \notag\\[-1mm]
&\mbox{C5: } {u_{k,r}}{M_{k,r}} \ge {u_{k,r}}{M_{\min }}, \forall k,r, \notag
\end{align}
\par
\vspace*{-3mm}
\noindent
where the user scheduling variable ${u_{k,r}}$ and antenna allocation variable ${M_{k,r}}$ are the optimization variables.
The objective function $\overline{R}_{\mathrm{sum}}$ denotes the conditional system sum-rate which can be given with ${R}_{\mathrm{sum}}$ in \eqref{SumRate} by substituting ${p_{k,r}} = \frac{p_{\mathrm{BS}}}{K}$ and ${\mathbf{G}} = {{\mathbf{I}}_{{N_{{\mathrm{RF}}}}}}$.
Similarly, ${\overline{R}_{k,r}}$ denotes the conditional individual data rate of user $k$ associated with RF chain $r$ which can be obtained with ${{R}_{k,r}}$ in \eqref{DLIndividualRate1} by substituting ${p_{k,r}} = \frac{p_{\mathrm{BS}}}{K}$ and ${\mathbf{G}} = {{\mathbf{I}}_{{N_{{\mathrm{RF}}}}}}$.
Constraint C2 restricts that each RF chain can only serve at most two users.
Constraint C3 is imposed such that one user can only be associated with at most one RF chain.
Constraint C4 guarantees that the number of all the allocated antennas on RF chain $r$ cannot be larger than $M_{\mathrm{BS}}$.
Constraint C5 is imposed to keep that the minimum number of antennas allocated to user $k$ is ${M_{\mathrm{min}}}$ if it is associated with RF chain $r$.
The reasons for introducing constraint C5 is two-fold: 1) we need to prevent the case where a user is not served by any antenna for a fair resource allocation; 2) constraint C5 can prevent the formation of high sidelobe beam, which introduces a high inter-group interference and might sacrifice the performance gain of multi-beam NOMA.
Note that constraint C5 are inactive when user $k$ is not assigned on RF chain $r$, i.e., $u_{k,r} = 0$.

The formulated problem is a non-convex integer programming problem, which is very difficult to find the globally optimal solution.
In particular, the effective channel gain in the objective function in \eqref{ResourceAllocation} involves a periodic trigonometric function of ${M_{k,r}}$, which cannot be solved efficiently with existing convex optimization methods\cite{Boyd2004}.
In general, the exhaustive search can optimally solve this problem, but its computational complexity is prohibitively high, which is given by
\begin{equation}\label{Complexity}
\mathcal{O}\left(({M_{{\mathrm{BS}}} - 2{M_{\min }}})^{{K - {N_{{\mathrm{RF}}}}}}\left( {\begin{array}{*{20}{c}}
K\\
{2{N_{RF}} - K}
\end{array}} \right)\mathop \Pi \limits_{r = 1}^{K - {N_{{\mathrm{RF}}}}} \left( {2r - 1} \right)\right),
\end{equation}
where $\mathcal{O}\left(\cdot\right)$ is the big-O notation.
As a compromise solution, we recast the formulated problem as a coalition formation game \cite{SaadCoalitionalGame,SaadCoalitional2012,WangCoalitionNOMA,SaadCoalitionOrder,Han2012} and propose a coalition formation algorithm for obtaining an efficient suboptimal solution of the user grouping and antenna allocation.

\subsubsection{The Proposed Coalition Formation Algorithm}
In this section, to derive an algorithm for the user grouping and antenna allocation with a low computational complexity, we borrow the concept from the coalitional game theory \cite{SaadCoalitionalGame,SaadCoalitional2012,WangCoalitionNOMA,SaadCoalitionOrder,Han2012} to achieve a suboptimal but effective solution.
Note that, although the game theory is commonly introduced to deal with the distributed resource allocation design problem\cite{SaadCoalitional2012}, it is also applicable to the scenario with a centralized utility \cite{SaadCoalitionOrder,WangCoalitionNOMA,Han2012}.
Besides, it is expected to achieve a better performance with centralized utility compared to the distributed resource allocation due to the availability of the overall resource allocation strategy and the system performance.

\textbf{Basic concepts:} We first introduce the notions from the coalitional game theory \cite{SaadCoalitionalGame,SaadCoalitional2012,WangCoalitionNOMA,SaadCoalitionOrder,Han2012} to reformulate the problem in \eqref{ResourceAllocation}.
In particular, we view all the $K$ users as a set of cooperative \emph{players}, denoted by $\mathcal{K} = \{1,\ldots,K\}$, who seek to form \emph{coalitions} $S$ (NOMA groups in this paper), $S \subseteq \mathcal{K} $, to maximize the conditional system sum-rate.
The \emph{coalition value}, denoted by ${V} \left(S\right)$, quantifies the payoff of a coalition in a game.
In this paper, we characterize the payoff of each player as its individual conditional data rate, since we aim to maximize the conditional system sum-rate in \eqref{ResourceAllocation}.
In addition, for our considered problem, since the payoff of each player in a coalition $S$ depends on the antenna allocation within the coalition and thus cannot be divided in any manner between the coalition members, our considered game has a nontransferable utility (NTU) property \cite{Han2012}.
Therefore, the coalition value is a set of payoff vectors, ${V} \left(S\right) \subseteq \mathbb{R}^{\left|S\right|}$, where each element represents the payoff of each player in $S$.
Furthermore, due to the existence of inter-group interference in our considered problem, the coalition value ${V} \left(S\right)$ depends not only on its own coalition members in $S$, but also on how the other players in $\mathcal{K}\backslash S$ are structured.
As a result, the considered game falls into the category of \emph{coalition game in partition form}, i.e., coalition formation game\cite{SaadCoalitionalGame}.

\textbf{Coalition value:} More specific, given a coalitional structure $\mathcal{B}$, defined as a partition of $\mathcal{K}$, i.e., a collection of coalitions $\mathcal{B} = \{S_1,\ldots,S_{\left|\mathcal{B}\right|}\}$, such that $\forall r \ne r'$, $S_{r} \bigcap S_{r'} = \emptyset$, and $\cup_{r = 1}^{\left|\mathcal{B}\right|} S_{r} = \mathcal{K}$, the value of a coalition $S_{r}$ is defined as ${V} \left(S_{r},\mathcal{B}\right)$.
Moreover, based on \eqref{DLIndividualRate1}, we can observe that ${\overline{R}_{k,r}}$ is affected by not only the antenna allocation strategy $\mathcal{M}_r = \{M_{k,r} | \forall k \in S_{r}\}$ in the coalition $S_{r}$, but also the antenna allocation strategy $\mathcal{M}_{r'} = \{M_{k,{r'}} | \forall k \in S_{{r'}}\}$ in other coalitions $S_{r'}$, $r' \neq r$.
Therefore, the coalition value is also a function of the antenna allocation strategy $\mathcal{M}_r$ of the coalition $S_{r}$ and the overall antenna allocation strategy $\Pi = \{\mathcal{M}_1,\ldots,\mathcal{M}_{\left|\mathcal{B}\right|}\}$, i.e., ${V} \left(S_{r},\mathcal{M}_r,\mathcal{B},\Pi\right)$.
Then, the coalition value of $S_{r}$ in place of $\mathcal{B}$ can be defined as
\vspace{-2mm}
\begin{equation}
{V} \left(S_{r},\mathcal{M}_r,\mathcal{B},\Pi\right) = \{v_k\left(S_{r},\mathcal{M}_r,\mathcal{B},\Pi\right) | \forall k \in S_{r} \},\vspace{-3mm}
\end{equation}
with the payoff of user $k$ as
\vspace{-2mm}
\begin{equation}\label{PayoffUserk}
v_k \left(S_{r},\mathcal{M}_r,\mathcal{B},\Pi\right) = \left\{ {\begin{array}{*{20}{c}}
\frac{{N_{\mathrm{RF}}}}{{\left|\mathcal{B}\right|}}{\overline{R}_{k,r}},&{{\left|\mathcal{B}\right|} > {N_{\mathrm{RF}}}},\\[-1mm]
{\overline{R}_{k,r}},&{{\left|\mathcal{B}\right|} \le {N_{\mathrm{RF}}}},
\end{array}} \right.\vspace{-2mm}
\end{equation}
where $\frac{{N_{\mathrm{RF}}}}{{\left|\mathcal{B}\right|}}$ is the time-sharing factor since the number of coalitions is larger than the number of system RF chains.
To facilitate the solution design, based on \eqref{DLIndividualRate1}, the conditional individual data rate of user $k$ associated with RF chain $r$, ${\overline{R}_{k,r}}$, can be rewritten with the notations in coalition game theory as follows
\vspace{-2mm}
\begin{equation}\label{DLIndividualRate1Game}
{\overline{R}_{k,r}} = \left\{ {\begin{array}{*{20}{c}}
{\log _2}\left( {1 + {\frac{{{p_{k,r}}{{\left| {{{\mathbf{\widetilde h}}}_k^{\mathrm{H}}}{{\mathbf{g}}_r} \right|}^2}}}{{\sum\limits_{r' \ne r, S_{r'} \in \mathcal{B}} {{{\left| {{{\mathbf{\widetilde h}}}_k^{\mathrm{H}}}{{\mathbf{g}}_{r'}} \right|}^2}\sum\limits_{d \in S_{r'}} {{p_{d,r}}}}  + \sum\limits_{d < k, d \in S_{r}} {{p_{d,r}}{{\left| {{{\mathbf{\widetilde h}}}_k^{\mathrm{H}}}{{\mathbf{g}}_r} \right|}^2}}  + {\sigma ^2}}}}} \right), &{k \in S_{r}},\\[-1mm]
0,&{\mathrm{otherwise}},
\end{array}} \right.\vspace{-2mm}
\end{equation}
with ${p_{k,r}} = \frac{p_{\mathrm{BS}}}{K}$ and ${\mathbf{G}} = {{\mathbf{I}}_{{N_{{\mathrm{RF}}}}}}$.

Based on the aforementioned coalitional game theory notions, the proposed coalition formation algorithm essentially forms coalitions among players iteratively to improve the system performance.
To facilitate the presentation of the proposed coalition formation algorithm, we introduce the following basic concepts.

\begin{Def}
Given two partitions $\mathcal{B}_1$ and $\mathcal{B}_2$ of the set of users $\mathcal{K}$ with two antenna allocation strategies $\Pi_1$ and $\Pi_2$, respectively, for two coalitions ${S_r} \in \mathcal{B}_1$ and ${S_{r'}} \in \mathcal{B}_2$ including user $k$, i.e., $k \in {S_r}$ and $k \in {S_{r'}}$, the \emph{preference relationship} for user $k \in \mathcal{K}$ is defined as
\vspace{-2mm}
\begin{equation}\label{PreferenceRelationship}
\left( {{S_r},\mathcal{B}_1} \right) { \preceq _k} \left( {{S_{r'}},\mathcal{B}_2} \right) \Leftrightarrow
\mathop {\max }\limits_{{\mathcal{M}_r}}   {\overline{R}_{{\mathrm{sum}}}}\left( {{\cal B}_1,\Pi_1} \right) \le \mathop {\max } \limits_{{\mathcal{M}_{r'}}} {\overline{R}_{{\mathrm{sum}}}} \left( {{\cal B}_2,\Pi_2} \right),\vspace{-2mm}
\end{equation}
with
\vspace{-2mm}
\begin{equation}\label{SystemSumRateCoalitionGame}
{\overline{R}_{{\mathrm{sum}}}} \left( {{\cal B},\Pi} \right) = \sum\limits_{{S_r} \in {\cal B}} {\sum\limits_{k \in {S_r}} {{v_k}\left( {{S_r},{\mathcal{M}_r},{\cal B},\Pi} \right)} }.\vspace{-2mm}
\end{equation}
The notation $\left( {{S_r},\mathcal{B}_1} \right){ \preceq _k} \left( {{S_{r'}},\mathcal{B}_2} \right)$ means that user $k$ prefers to be part of coalition ${S_{r'}}$ when $\mathcal{B}_2$ is in place, over being part of coalition ${S_{r}}$ when $\mathcal{B}_1$ is in place, or at least prefers both pairs of coalitions and partitions equally.
As shown in \eqref{PreferenceRelationship} and \eqref{SystemSumRateCoalitionGame}, we have $\left( {{S_r},\mathcal{B}_1} \right){ \preceq _k} \left( {{S_{r'}},\mathcal{B}_2} \right)$ if and only if the resulting conditional system sum-rate with $\mathcal{B}_2$ and $\Pi_2$ is larger than or at least equal to that with $\mathcal{B}_1$ and $\Pi_1$, with the optimized antenna allocation over ${\mathcal{M}_r}$ and ${\mathcal{M}_{r'}}$, respectively.
Furthermore, we denote their asymmetric counterpart ${ \prec_k}$ and $<$ to indicate the \emph{strictly preference relationship}.
\end{Def}

Note that, the maximization $\mathop {\max }  \nolimits_{{\mathcal{M}_r}}   {\overline{R}_{{\mathrm{sum}}}}\left( {{\cal B}_1,\Pi_1} \right)$ in \eqref{PreferenceRelationship} is only over ${\mathcal{M}_r}$ for the involved coalition ${S_r}$ via fixing all the other antenna allocation strategies in $\{\Pi_1 \backslash {\mathcal{M}_r}\}$ when $\mathcal{B}_1$ is in place.
Similarly, the maximization $\mathop {\max }  \nolimits_{{\mathcal{M}_{r'}}}   {\overline{R}_{{\mathrm{sum}}}}\left( {{\cal B}_2,\Pi_2} \right)$ in \eqref{PreferenceRelationship} is only over ${\mathcal{M}_{r'}}$ for the involved coalition ${S_{r'}}$ via fixing all the other antenna allocation strategies in $\{\Pi_2 \backslash {\mathcal{M}_{r'}}\}$ when $\mathcal{B}_2$ is in place.
Let ${\mathcal{M}_r^{*}} = \mathop {\arg } \max \nolimits_{{\mathcal{M}_r}}   {\overline{R}_{{\mathrm{sum}}}}\left( {{\cal B}_1,\Pi_1} \right)$ and ${\mathcal{M}_{r'}^{*}} = \mathop {\arg } \max\nolimits_{{\mathcal{M}_{r'}}} {\overline{R}_{{\mathrm{sum}}}} \left( {{\cal B}_2,\Pi_2} \right)$ denote the optimal solutions for the maximization problems in both sides of  \eqref{PreferenceRelationship}.
To find the optimal ${\mathcal{M}_r^{*}}$ and ${\mathcal{M}_{r'}^{*}}$ in \eqref{PreferenceRelationship}, we use the full search method over ${\mathcal{M}_r}$ and ${\mathcal{M}_{r'}}$ within the feasible set of C4 and C5 in \eqref{ResourceAllocation}.
Recall that there are at most two members in each coalition, the computational complexity for solving the antenna allocation in each maximization problem is acceptable via a one-dimensional full search over the integers in set $[{M_{\min }},{M_{{\mathrm{BS}}} - {M_{\min }}}]$.
Based on the defined preference relationship above, we can define the \emph{strictly preferred leaving and joining operation} in the $\mathrm{iter}$-th iteration as follows.
\begin{Def}\label{LeaveAndJoin}
In the $\mathrm{iter}$-th iteration, given the current partition $\mathcal{B}_{\mathrm{iter}} = \{S_1,\ldots,S_{\left|\mathcal{B}_{\mathrm{iter}}\right|}\}$ and its corresponding antenna allocation strategy $\Pi_{\mathrm{iter}} = \{\mathcal{M}_1,\ldots,\mathcal{M}_{\left|\mathcal{B}_{\mathrm{iter}}\right|}\}$ of the set of users $\mathcal{K}$, user $k$ will \emph{leave} its current coalition ${S_{r}}$ and \emph{joint} another coalition ${S_{r'}} \in \mathcal{B}_{\mathrm{iter}}$, $r' \neq r$, if and only if it is a \emph{strictly preferred leaving and joining operation}, i.e.,
\vspace{-2mm}
\begin{equation}\label{StrictPreferredLaJ}
\left( {{S_r},\mathcal{B}_{\mathrm{iter}}} \right) { \prec _k} \left( {{S_{r'}}  \cup  \{k\} ,\mathcal{B}_{\mathrm{iter}+1}} \right),\vspace{-2mm}
\end{equation}
where new formed partition is $\mathcal{B}_{\mathrm{iter}+1} = \left\{ {\mathcal{B}_{\mathrm{iter}}\backslash \left\{ {{S_r},{S_{r'}}} \right\}} \right\} \cup \left\{ {{S_r}\backslash \left\{ k \right\},{S_{r'}} \cup \left\{ k \right\}} \right\}$.
Let ${{\cal M}_r^\circ}$ denote the optimal antenna allocation strategy for the coalition $\{{S_r}\backslash \left\{ k \right\}\}$ if user $k$ leaves the coalition ${S_{r}}$.
Due to ${\left|S_r\right|} \le 2$, there are at most one member left in $\{{S_r}\backslash \left\{ k \right\}\}$, and thus we have
\vspace{-3mm}
\begin{equation}
{{\cal M}_r^\circ} = \left\{ {\begin{array}{*{20}{c}}
M_{\mathrm{BS}}, & \left|{S_r}\backslash \left\{ k \right\}\right| = 1,\\[-1mm]
0,           & \left|{S_r}\backslash \left\{ k \right\}\right| = 0.
\end{array}} \right.
\end{equation}
On the other hand, we denote ${\cal M}_{r'}^*$ as the optimal antenna allocation to maximize the conditional system sum-rate for the new formed coalition ${S_{r'}} \cup \left\{ k \right\}$ if user $k$ joins the coalition ${S_{r'}}$.
Similarly, we note that the maximization is only over ${\cal M}_{r'}$ via fixing all the others antenna strategies $\left\{ {\Pi_{\mathrm{iter}}\backslash \left\{ {{\cal M}_r,{{\cal M}_{r'}}} \right\}} \right\} \cup \left\{ {{{\cal M}_r^\circ} } \right\}$.
Then, we can define the new antenna allocation strategy for the new formed partition $\mathcal{B}_{\mathrm{iter}+1}$ as $\Pi_{\mathrm{iter}+1} = \left\{ {\Pi_{\mathrm{iter}}\backslash \left\{ {{\cal M}_r,{{\cal M}_{r'}}} \right\}} \right\} \cup \left\{ {{{\cal M}_r^\circ}, {{\cal M}_{r'}^*} } \right\}$.
In other words, we have the update rule as $\left\{ {{S_r},{S_{r'}}} \right\} \to \left\{ {{S_r}\backslash \left\{ k \right\},{S_{r'}} \cup \left\{ k \right\}} \right\}$, $\left\{ {{\cal M}_r,{{\cal M}_{r'}}} \right\}\to \left\{ {{{\cal M}_r^\circ},{\cal M}_{r'}^* } \right\}$, $\mathcal{B}_{\mathrm{iter}} \to \mathcal{B}_{\mathrm{iter}+1}$, and $\Pi_{\mathrm{iter}} \to \Pi_{\mathrm{iter}+1}$.
\end{Def}

The defined \emph{strictly preferred leaving and joining operation} above provides a mechanism to decide whether user $k$ should move from ${S_r}$ to ${S_{r'}}$, given that the coalition and partition pair $\left( {{S_{r'}}  \cup  \{k\} ,\mathcal{B}_{\mathrm{iter}+1}} \right)$ is strictly preferred over $\left( {{S_r},\mathcal{B}_{\mathrm{iter}}} \right)$.
However, as we mentioned before, there is a constraint on the size of coalition, i.e., ${\left|S_r\right|} \le 2$, $\forall r$.
As a result, we should prevent the size of the new coalition being larger than 2, i.e., $\left|\{{S_{r'}} \cup \left\{ k \right\}\}\right| > 2$.
To this end, we need to introduce the concept of \emph{strictly preferred switch operation} \cite{SaadCoalitional2012} as follows to enable user $k \in {S_{r}}$ and user $k' \in {S_{r'}}$ switch with each other, such that the new formed coalitions satisfy $\left| \{{S_{r}} \backslash \{k\} \cup \left\{ k' \right\}\} \right| \le 2$ and $\left| \{{S_{r'}} \backslash \{k'\} \cup \left\{ k \right\}\} \right| \le 2$.

\begin{Def}\label{SwitchOperation}
In the $\mathrm{iter}$-th iteration, given a partition $\mathcal{B}_{\mathrm{iter}} = \{S_1,\ldots,S_{\left|\mathcal{B}_{\mathrm{iter}}\right|}\}$ and an corresponding antenna allocation strategy $\Pi_{\mathrm{iter}} = \{\mathcal{M}_1,\ldots,\mathcal{M}_{\left|\mathcal{B}_{\mathrm{iter}}\right|}\}$ of the set of users $\mathcal{K}$, user $k \in {S_{r}}$ and user $k' \in {S_{r'}}$ will switch with each other, if and only if it is a \emph{strictly preferred switch operation}, i.e.,
\vspace{-2mm}
\begin{align}\label{SwitchRule}
\left( {{S_r},{S_r'},\mathcal{B}_{\mathrm{iter}}} \right) &{ \prec ^{k'}_k} \left({S_{r}} \backslash \{k\} \cup \left\{ k' \right\}, {{S_{r'}} \backslash \{k'\} \cup \left\{ k \right\} ,\mathcal{B}_{\mathrm{iter}+1}} \right) \notag\\[-2mm]
\Leftrightarrow \mathop {\max }\limits_{{\mathcal{M}_r},{\mathcal{M}_r'}}  {\overline{R}_{{\mathrm{sum}}}}\left( {{\cal B}_{\mathrm{iter}},\Pi_{\mathrm{iter}}} \right)  &< \mathop {\max } \limits_{{\mathcal{M}_r},{\mathcal{M}_r'}} {\overline{R}_{{\mathrm{sum}}}}\left( {{\cal B}_{\mathrm{iter}+1},\Pi_{\mathrm{iter}+1}} \right),
\end{align}
where new formed partition is $\mathcal{B}_{\mathrm{iter}+1} = \left\{ {{\cal B}_{\mathrm{iter}}\backslash \left\{ {{S_r},{S_{r'}}} \right\}} \right\} \cup \left\{ {\{{S_{r}} \backslash \{k\} \cup \left\{ k' \right\}\},\{{S_{r'}} \backslash \{k'\} \cup \left\{ k \right\}\}} \right\}$.
Let ${\cal M}_r^ +$ and ${\cal M}_{r'}^ +$ denote the optimal solutions for the maximization on the right hand side of \eqref{SwitchRule}.
Then, we can define the new antenna allocation strategy for the new formed partition $\mathcal{B}_{\mathrm{iter}+1}$ as ${\Pi_{\mathrm{iter}+1} } = \left\{ {\Pi_{\mathrm{iter}}\backslash \left\{ {{\cal M}_r,{\cal M}_{r'}} \right\}} \right\} \cup \left\{ {{\cal M}_r^ + ,{\cal M}_{r'}^ + } \right\}$.
In other words, we have the update rule $\left\{ {{S_r},{S_{r'}}} \right\} \to \left\{ {{S_{r}} \backslash \{k\} \cup \left\{ k' \right\},{S_{r'}} \backslash \{k'\} \cup \left\{ k \right\}} \right\}$, $\left\{ {{\cal M}_r,{{\cal M}_{r'}}} \right\}\to \left\{ {{{\cal M}_r^{+}},{\cal M}_{r'}^{+} } \right\}$, $\mathcal{B}_{\mathrm{iter}} \to \mathcal{B}_{\mathrm{iter}+1}$, and $\Pi_{\mathrm{iter}} \to \Pi_{\mathrm{iter}+1}$.
\end{Def}

Again, the maximization in \eqref{SwitchRule} is only over ${\cal M}_r$ and ${\cal M}_{r'}$ within the feasible set of C4 and C5 in \eqref{ResourceAllocation} for the involved coalitions ${S_{r}}$ and ${S_{r'}}$ in the switch operation, via fixing all the other antenna allocation strategies $\left\{ {\Pi_{\mathrm{iter}}\backslash \left\{ {{\cal M}_r,{\cal M}_{r'}} \right\}} \right\}$.
Also, the computational complexity for each maximization is acceptable via a two-dimensional full search over the integers in $[{M_{\min }},{M_{{\mathrm{BS}}} - {M_{\min }}}]$.
From \eqref{SwitchRule}, we can observe that a switch operation is a strictly preferred switch operation of users $k$ and $k'$ if and only if when this switch operation can strictly increase the conditional system sum-rate.
The defined \emph{strictly preferred switch operation} above provides a mechanism to decide whether to switch user $k$ and user $k'$ with each other, given that it is a strictly preferred switch operation by users $k$ and $k'$.

\begin{table}
\vspace{-12mm}
\begin{algorithm} [H]                    
\caption{User Grouping and Antenna Allocation Algorithm}
\label{alg1}                             
\begin{algorithmic} [1]
\footnotesize          
\STATE \textbf{Initialization}\\
Initialize the iteration index $\mathrm{iter} = 0$.
The partition is initialized by $\mathcal{B}_0 = \mathcal{K} = \{S_1,\ldots,S_{K}\}$ with $S_k = {k}$, $\forall k$, i.e., OMA.
Correspondingly, the antenna allocation is initialized with $\Pi_0 = \{\mathcal{M}_1,\ldots,\mathcal{M}_K\}$ with $\mathcal{M}_k = \{{M_{\mathrm{BS}}}\}$, $\forall k$.
\REPEAT
\FOR{$k$ = $1$:$K$}
    \STATE User $k \in {S_{r}}$ visits each existing coalitions ${S_{r'}} \in \mathcal{B}_{\mathrm{iter}}$ with ${S_{r'}} \neq {S_{r}}$.
    \IF {$\left|\{{S_{r'}} \cup \left\{ k \right\}\}\right| <= 2$}
        \IF {$\left( {{S_r},\mathcal{B}_{\mathrm{iter}}} \right) { \prec _k} \left( {{S_{r'}}  \cup  \{k\} ,\mathcal{B}_{\mathrm{iter}+1}} \right)$}
            \STATE Execute the leaving and joining operation in Definition \ref{LeaveAndJoin}.
            \STATE $\mathrm{iter} = \mathrm{iter} + 1$.
        \ENDIF
    \ELSE
        \IF {$\left( {{S_r},{S_r'},\mathcal{B}_{\mathrm{iter}}} \right) { \prec ^{k'}_k} \left({S_{r}} \backslash \{k\} \cup \left\{ k' \right\}, {{S_{r'}} \backslash \{k'\} \cup \left\{ k \right\} ,\mathcal{B}_{\mathrm{iter}+1}} \right)$}
            \STATE Execute the switch operation in Definition \ref{SwitchOperation}.
            \STATE $\mathrm{iter} = \mathrm{iter} + 1$.
        \ENDIF
    \ENDIF
\ENDFOR
\UNTIL {No strictly preferred leaving and joining operation or switch operation can be found.}
\RETURN {$\mathcal{B}_{\mathrm{iter}}$ and $\Pi_{\mathrm{iter}}$}
\end{algorithmic}
\end{algorithm}
\vspace{-17mm}
\end{table}

Now, the proposed coalition algorithm for the user grouping and antenna allocation problem in \eqref{ResourceAllocation} is shown in \textbf{Algorithm} \ref{alg1}.
The algorithm is initialized with each user as a coalition, i.e., OMA, and each user is allocated with the whole antenna array.
In each iteration, all the users visit all the potential coalitions except its own coalition in current coalitional structure, i.e., ${S_{r'}} \in \mathcal{B}_{\mathrm{iter}}$ and ${S_{r'}} \neq {S_{r}}$.
Then, each user checks and executes the leaving and joining operation or the switch operation based on the Definition \ref{LeaveAndJoin} and Definition \ref{SwitchOperation}, respectively.
The iteration stops when no more preferred operation can be found.

\subsubsection{Effectiveness, Stability, and Convergence}
In the following, we briefly discuss the effectiveness, stability, and convergence for our proposed coalition formation algorithm due to the page limit.
Interested readers are referred to \cite{SaadCoalitional2012,WangCoalitionNOMA,SaadCoalitionOrder} for detailed proofs.
From the Definition \ref{LeaveAndJoin} and Definition \ref{SwitchOperation}, we can observe that every executed operation in \textbf{Algorithm} \ref{alg1} increases the conditional system sum-rate.
In other words, \textbf{Algorithm} \ref{alg1} can effectively increase the conditional system sum-rate in \eqref{ResourceAllocation}.
The stability of \textbf{Algorithm} \ref{alg1} can be proved by contradiction, cf. \cite{SaadCoalitional2012,WangCoalitionNOMA,SaadCoalitionOrder}.
Firstly, we need to note that each user has an incentive to leave its current coalition only and if only when this operation can increase the conditional system sum-rate, i.e., a strictly preferred leaving and joining operation or switch operation.
Then, we can define a stable coalitional structure state as a final state $\mathcal{B}^{*}$ after \textbf{Algorithm} \ref{alg1} terminates where no user has an incentive to leave its current coalition.
If there exists a user $k$ want to leave its current coalition $S_r$ in the final coalitional structure $\mathcal{B}^{*}$, it means that \textbf{Algorithm} \ref{alg1} will not terminate and $\mathcal{B}^{*}$ is not a final state, which causes a contradiction.

Now, the convergence of \textbf{Algorithm} \ref{alg1} can be proved with the following logic.
Since the number of feasible combinations of user grouping and antenna allocation in \eqref{ResourceAllocation} is finite, the number of strictly preferred operations is finite.
Moreover, according to \textbf{Algorithm} \ref{alg1}, the conditional system sum-rate increases after each approved operation.
Since the conditional system sum-rate is upper bounded by above due to the limited number of RF chains and time resources, \textbf{Algorithm} \ref{alg1} terminates when the conditional system sum-rate is saturated.
In other words, \textbf{Algorithm} \ref{alg1} converges to the final stable coalitional structure $\mathcal{B}^{*}$ within a limited number of iterations.

\subsubsection{Computational Complexity}
Assume that, in the $\mathrm{iter}$-th iteration,  a coalitional structure $\mathcal{B}_{\mathrm{iter}}$ consists of $\left|\mathcal{B}^{\mathrm{I}}_{\mathrm{iter}}\right|$ single-user coalitions and $\left|\mathcal{B}^{\mathrm{II}}_{\mathrm{iter}}\right|$ two-users coalitions with $\left|\mathcal{B}_{\mathrm{iter}}\right| = \left|\mathcal{B}^{\mathrm{I}}_{\mathrm{iter}}\right| + \left|\mathcal{B}^{\mathrm{II}}_{\mathrm{iter}}\right|$.
For user $k$, the computational complexity to locate a strictly preferred leaving and joining operation is $\mathcal{O}\left(2(\left|\mathcal{B}^{\mathrm{I}}_{\mathrm{iter}}\right| + 1) ({M_{{\mathrm{BS}}} - 2{M_{\min }}})\right)$ in the worst case and the counterpart to locate a strictly preferred switch operation is $\mathcal{O}\left(4\left|\mathcal{B}^{\mathrm{II}}_{\mathrm{iter}}\right| ({M_{{\mathrm{BS}}} - 2{M_{\min }}})^2\right)$ in the worst case.
As a result, the computational complexity in each iteration of our proposed coalition formation algorithm is $\mathcal{O}\left( 2(\left|\mathcal{B}^{\mathrm{I}}_{\mathrm{iter}}\right| + 1) ({M_{{\mathrm{BS}}} - 2{M_{\min }}})  + 4\left|\mathcal{B}^{\mathrm{II}}_{\mathrm{iter}}\right| ({M_{{\mathrm{BS}}} - 2{M_{\min }}})^2\right)$ in the worst case, which is substantially low compared to that of the exhaustive search. i.e., \eqref{Complexity}.

\subsection{Second Stage: Digital Precoder and Power Allocation Design}
\subsubsection{ZF Digital Precoder}
Given the obtained user grouping strategy $\mathcal{B}^{*} = \{S_1^{*},\ldots,S_{N_{\mathrm{RF}}}^{*}\}$ and the antenna allocation strategy $\Pi^{*}$ in the first stage, we can obtain the effective channel matrix ${{\mathbf{\widetilde H}}} \in \mathbb{C}^{{N_{\mathrm{RF}} \times K}}$ via uplink pilot transmission.
We adopt a ZF digital precoder to suppress the inter-group interference.
Since there might be more than one users in each group $S_r^{*} \in \mathcal{B}^{*}$, we perform singular value decomposition (SVD) on the equivalent channel for each NOMA group $S_r^{*}$ with $\left|S_r^{*}\right| = 2$.
In particular, let ${{\mathbf{\widetilde H}}}_r \in \mathbb{C}^{{N_{\mathrm{RF}} \times \left|S_r^{*}\right|}}$, $\forall r$, denotes the effective channel matrix for all the $\left|S_r^{*}\right|$ users in the coalition $S_r^{*}$.
We have the SVD for ${{\mathbf{\widetilde H}}}_r$ as follows:
\vspace{-2mm}
\begin{equation}
{{\mathbf{\widetilde H}}}_r^{\mathrm{H}} = {\mathbf{U}_r}{\mathbf{\Sigma} _r}\mathbf{V}_r^{\mathrm{H}},\vspace{-2mm}
\end{equation}
where ${\mathbf{U}_r} = {\left[{{\bf{u}}_{r,{1}}}, \cdots ,{{\bf{u}}_{r,{\left|S_r\right|}}}\right]} \in \mathbb{C}^{{ \left|S_r^{*}\right| \times \left|S_r^{*}\right|}}$ is the left singular matrix, ${\mathbf{\Sigma} _r} \in \mathbb{R}^{{ \left|S_r^{*}\right| \times N_{\mathrm{RF}}}}$ is the singular value matrix with its diagonal entries as singular values in descending order, and ${\mathbf{V}_r} \in \mathbb{C}^{{ N_{\mathrm{RF}} \times N_{\mathrm{RF}}}}$ is the right singular matrix.
Then, the equivalent channel vector of the NOMA group $S_r^{*}$ is given by
\vspace{-2mm}
\begin{equation}
{{\mathbf{\hat h}}}_r = {{\mathbf{\widetilde H}}}_r {{\bf{u}}_{r,{1}}} \in \mathbb{C}^{{N_{\mathrm{RF}} \times 1}},\vspace{-2mm}
\end{equation}
where ${{\bf{u}}_{r,{1}}} \in \mathbb{C}^{{ \left|S_r^{*}\right| \times 1}}$ is the first left singular vector corresponding to the maximum singular value.
Note that, the equivalent channel for a single-user coalition with $\left|S_r^{*}\right| = 1$ can be directly given with its effective channel, i.e., ${{\mathbf{\hat h}}}_r = {{\mathbf{\widetilde h}}}_r$.
Now, the equivalent channel for all the coalitions on all the RF chains can be given by
\vspace{-2mm}
\begin{equation}
{\bf{\hat H}} = {\left[ {
{{\bf{\hat{h}}}_1}, \ldots,{{\bf{\hat{h}}}_{N_{\mathrm{RF}}}}
} \right]} \in \mathbb{C}^{{ {N_{\mathrm{RF}}} \times N_{\mathrm{RF}}}}.\vspace{-2mm}
\end{equation}
Furthermore, the ZF digital precoder can be obtained by
\vspace{-2mm}
\begin{equation}
{\mathbf{G}} = {\bf{\hat H}}^{\mathrm{H}}\left({\bf{\hat H}}{\bf{\hat H}}^{\mathrm{H}}\right)^{-1} \in \mathbb{C}^{{ N_{\mathrm{RF}} \times {N_{\mathrm{RF}}} }},\vspace{-2mm}
\end{equation}
where ${\mathbf{G}} = \left[ {
{{{\mathbf{g}}_1}},\ldots ,{{{\mathbf{g}}_{{N_{\mathrm{RF}}}}}}
} \right]$ and ${{{\mathbf{g}}_r}}$ denotes the digital precoder shared by all the user in $S_r^{*}$.

\subsubsection{Power Allocation Design}
Given the effective channel matrix ${{\mathbf{\widetilde H}}}_r$ and the digital precoder ${\mathbf{G}}$, the optimal power allocation can be formulated as the following optimization problem:
\vspace{-4mm}
\begin{align} \label{ResourceAllocation_Power}
&\underset{{p_{k,r}}}{\maxo} \;\;R_{\mathrm{sum}}\notag\\[-2mm]
\mbox{s.t.}\;\;
&\mbox{C1: } {p_{k,r}} \ge 0, \forall k,r,
\;\;\mbox{C2: } \sum\nolimits_{k = 1}^K\sum\nolimits_{r = 1}^{N_{{\mathrm{RF}}}} {{u_{k,r}^{*}}{p_{k,r}}}  \le {p_{{\mathrm{BS}}}}, \notag\\[-1mm]
&\mbox{C3: } {u_{k,r}^{*}}R_{k,i,r} \ge {u_{k,r}^{*}}R_{i,r}, \forall i > k, \forall r,
\;\;\mbox{C4: } \sum\nolimits_{r = 1}^{{N_{{\mathrm{RF}}}}} R_{k,r} \ge R_{\mathrm{min}}, \forall k,
\end{align}
\par
\vspace*{-2mm}
\noindent
where $R_{k,r}$, $R_{k,i,r}$, and $R_{\mathrm{sum}}$ are given by \eqref{DLIndividualRate1}, \eqref{IndividualRate2}, and \eqref{SumRate} with replacing ${u_{k,r}}$ with ${u_{k,r}^{*}}$, respectively.
Note that the user scheduling ${u_{k,r}^{*}}$ can be easily obtained by the following mapping:
\vspace{-4mm}
\begin{equation}
{u_{k,r}^{*}} = \left\{ {\begin{array}{*{20}{c}}
1,&{{\mathrm{if}}\;k \in S_r^{*}},\\[-1mm]
0,&{{\mathrm{otherwise}}}.
\end{array}} \right.\vspace{-1mm}
\end{equation}
Constraint C2 is the total power constraint at the BS.
Constraint C3 is introduced to guarantee the success of SIC decoding.
Note that constraint C3 are inactive when $u_{k,r}^{*} = 0$ or $u_{i,r}^{*} = 0$.
Constraint C4 is imposed to guarantee a minimum rate requirement for each user.

The formulated problem is a non-convex optimization, but can be equivalently transformed to a canonical form of D.C. programming \cite{WeiTCOM2017} as follows:
\vspace{-2mm}
\begin{align} \label{ResourceAllocation_Power1}
&\underset{{p_{k,r}}}{\mino} \;\;{{H_1}\left( {\bf{p}} \right) - {H_2}\left( {\bf{p}} \right)}\notag\\[-2mm]
\mbox{s.t.}\;\;
&\mbox{C1: } {p_{k,r}} \ge 0, \forall k,r,
\;\;\mbox{C2: } \sum\nolimits_{k = 1}^K\sum\nolimits_{r = 1}^{N_{{\mathrm{RF}}}}  {{u_{k,r}^{*}}{p_{k,r}}}  \le {p_{{\mathrm{BS}}}}, \notag\\[-1mm]
&\mbox{C3: } u_{k,r}^*u_{i,r}^*{{{{\left| {{{{\bf{\widetilde h}}}_i^{\mathrm{H}}}{{\bf{g}}_r}} \right|}^2}}} D_2^{k,i,r}\left( {\bf{p}} \right) \le u_{k,r}^*u_{i,r}^*{{{{\left| {{{{\bf{\widetilde h}}}_k^{\mathrm{H}}}{{\bf{g}}_r}} \right|}^2}}}D_2^{i,i,r}\left( {\bf{p}} \right), \forall i > k, \forall r,\notag\\[-1mm]
&\mbox{C4: } {u_{k,r}^*{p_{k,r}}{{{{\left| {{{{\bf{\widetilde h}}}_k^{\mathrm{H}}}{{\bf{g}}_r}} \right|}^2}}} \ge \left( {{2^{u_{k,r}^*{R_{{\mathrm{min}}}}}} - 1} \right)D_2^{k,k,r}\left( {\bf{p}} \right)}, \forall k,
\end{align}
\par
\vspace*{-2mm}
\noindent
where ${\bf{p}} \in \mathbb{R}^{{ K {N_{\mathrm{RF}}}} \times 1}$ denotes the collection of ${p_{k,r}}$, ${H_1}\left( {\bf{p}} \right)$ and ${H_2}\left( {\bf{p}} \right)$ are given by
\vspace*{-1mm}
\begin{equation}
\hspace{-2mm}{H_1}\hspace{-1mm}\left( {\bf{p}} \right)\hspace{-1mm} =\hspace{-1mm} -\hspace{-1mm}\sum\nolimits_{k = 1}^K {\sum\nolimits_{r = 1}^{{N_{RF}}} {{{\log }_2}\hspace{-1mm}\left( D_1^{k,k,r}\hspace{-1mm}\left( {\bf{p}} \right)\hspace{-1mm} \right)\hspace{-1mm}} } \;\;\mathrm{and} \;\;
{H_2}\hspace{-1mm}\left( {\bf{p}} \right)\hspace{-1mm} = \hspace{-1mm}-\hspace{-1mm}\sum\nolimits_{k = 1}^K {\sum\nolimits_{r = 1}^{{N_{RF}}} {{{\log }_2}\hspace{-1mm}\left( D_2^{k,k,r}\hspace{-1mm}\left( {\bf{p}}\right)\hspace{-1mm} \right)} },\vspace*{-1mm}
\end{equation}
respectively, and $D_1^{k,i,r}\left( {\bf{p}} \right)$ and $D_2^{k,i,r}\left( {\bf{p}} \right)$ are given by
\vspace*{-1mm}
\begin{align}
D_1^{k,i,r}\left( {\bf{p}} \right) &= \sum\nolimits_{r' \ne r}^{{N_{{\mathrm{RF}}}}} {{{\left| {{{{\bf{\widetilde h}}}_k^{\mathrm{H}}}{{\bf{g}}_{r'}}} \right|}^2}\sum\nolimits_{d = 1}^K {u_{d,r'}^*{p_{d,r'}}} }  + \sum\nolimits_{d = 1}^i {u_{d,r}^*{p_{d,r}}} {\left| {{{{\bf{\widetilde h}}}_k^{\mathrm{H}}}{{\bf{g}}_r}} \right|^2} + {\sigma ^2} \;\mathrm{and} \notag\\[-1mm]
D_2^{k,i,r}\left( {\bf{p}} \right) &= \sum\nolimits_{r' \ne r}^{{N_{{\mathrm{RF}}}}} {{{\left| {{{{\bf{\widetilde h}}}_k^{\mathrm{H}}}{{\bf{g}}_{r'}}} \right|}^2}\sum\nolimits_{d = 1}^K {u_{d,r'}^*{p_{d,r'}}} }  + \sum\nolimits_{d = 1}^{i - 1} {u_{d,r}^*{p_{d,r}}} {\left| {{{{\bf{\widetilde h}}}_k^{\mathrm{H}}}{{\bf{g}}_r}} \right|^2} + {\sigma ^2},
\end{align}
\par
\vspace*{-1mm}
\noindent
respectively.
Note that ${H_1}\left( {\bf{p}} \right)$ and ${H_2}\left( {\bf{p}} \right)$ are differentiable convex functions with respect to ${\bf{p}}$.
Thus, for any feasible solution ${\bf{p}}^{\mathrm{iter}}$ in the $\mathrm{iter}$-th iteration, we can obtain a lower bound for ${H_2}\left( {\bf{p}} \right)$, which is given by
\vspace{-2mm}
\begin{equation}
{H_2}\left( {\bf{p}} \right) \ge {H_2}\left( {\bf{p}}^{\mathrm{iter}} \right) + {\nabla _{\bf{p}}}{H_2}{\left( {\bf{p}}^{\mathrm{iter}} \right)^{\mathrm{T}}}\left( {\bf{p}} - {\bf{p}}^{\mathrm{iter}} \right),\vspace{-1mm}
\end{equation}
with ${\nabla _{\bf{p}}}{H_2}\left( {{{\bf{p}}^{{\mathrm{iter}}}}} \right) = \left\{ {\frac{{\partial {H_2}\left( {\bf{p}} \right)}}{{\partial {p_{k,r}}}}\left| {_{{{\bf{p}}^{{\mathrm{iter}}}}}} \right.} \right\}_{k = 1,r = 1}^{k = K,r = {N_{{\mathrm{RF}}}}} \in \mathbb{R}^{{ K {N_{\mathrm{RF}}}} \times 1}$ denoting the gradient of ${H_2}\left( \cdot\right)$ with respect to ${\bf{p}}$ and
\begin{equation}
\hspace{-2mm}{\frac{{\partial {H_2}\hspace{-1mm}\left( {\bf{p}} \right)\hspace{-1mm}}}{{\partial {p_{k,r}}}}\left| {_{{{\bf{p}}^{{\mathrm{iter}}}}}} \right.} = -\frac{1}{\log(2)} \sum\limits_{k' = 1}^K {\sum\limits_{r' \ne r}^{{N_{RF}}} {\frac{{{{\left| {{{{\bf{\widetilde h}}}_{k'}^{\mathrm{H}}}{{\bf{g}}_r}} \right|}^2}u_{k,r}^*}}{D_2^{k',k',r'}\left( {{{\bf{p}}^{{\mathrm{iter}}}}} \right)}} }
 - \frac{1}{\log(2)} \sum\limits_{k' = k + 1}^K {\frac{{{{\left| {{{{\bf{\widetilde h}}}_{k'}^{\mathrm{H}}}{{\bf{g}}_r}} \right|}^2}u_{k,r}^*}}{{D_2^{k',k',r}\left( {{{\bf{p}}^{{\mathrm{iter}}}}} \right)}}}.\vspace*{-1mm}
\end{equation}
Then, we obtain an upper bound for the minimization problem in \eqref{ResourceAllocation_Power1} by solving the following convex optimization problem:
\vspace{-2mm}
\begin{align} \label{ResourceAllocation_Power2}
&\underset{{p_{k,r}}}{\mino} \;\;{{H_1}\left( {\bf{p}} \right) - {H_2}\left( {\bf{p}}^{\mathrm{iter}} \right) - {\nabla _{\bf{p}}}{H_2}{\left( {\bf{p}}^{\mathrm{iter}} \right)^{\mathrm{T}}}\left( {\bf{p}} - {\bf{p}}^{\mathrm{iter}} \right)}\notag\\[-2mm]
\mbox{s.t.}\;\;
&\mbox{C1: } {p_{k,r}} \ge 0, \forall k,r,
\;\;\mbox{C2: } \sum\nolimits_{k = 1}^K \sum\nolimits_{r = 1}^{N_{{\mathrm{RF}}}}  {{u_{k,r}^{*}}{p_{k,r}}}  \le {p_{{\mathrm{BS}}}}, \forall r, \notag\\[-1mm]
&\mbox{C3: } u_{k,r}^*u_{i,r}^*{{{{\left| {{{{\bf{\widetilde h}}}_i^{\mathrm{H}}}{{\bf{g}}_r}} \right|}^2}}} D_2^{k,i,r}\left( {\bf{p}} \right) \le u_{k,r}^*u_{i,r}^*{{{{\left| {{{{\bf{\widetilde h}}}_k^{\mathrm{H}}}{{\bf{g}}_r}} \right|}^2}}}D_2^{i,i,r}\left( {\bf{p}} \right), \forall i > k, \notag\\[-1mm]
&\mbox{C4: } {u_{k,r}^*{p_{k,r}}{{{{\left| {{{{\bf{\widetilde h}}}_k^{\mathrm{H}}}{{\bf{g}}_r}} \right|}^2}}} \ge \left( {{2^{u_{k,r}^*{R_{{\mathrm{min}}}}}} - 1} \right)D_2^{k,k,r}\left( {\bf{p}} \right)}, \forall k.
\end{align}
\par
\vspace*{-2mm}
\noindent

\begin{table}
\vspace{-12mm}
\begin{algorithm} [H]                    
\caption{Power Allocation Algorithm}     
\label{alg2}                             
\begin{algorithmic} [1]
\footnotesize          
\STATE \textbf{Initialization}\\
Initialize the convergence tolerance $\epsilon$, the maximum number of iterations $\mathrm{iter}_\mathrm{max}$, the iteration index $\mathrm{iter} = 1$, and the initial feasible solution $\mathbf{{p}}^{\mathrm{iter}}$.

\REPEAT
\STATE Solve \eqref{ResourceAllocation_Power2} for a given $\mathbf{{p}}^{\mathrm{iter}}$ to obtain the power allocation $\mathbf{{p}}^{\mathrm{iter}+1}$.
\STATE Set $\mathrm{iter}=\mathrm{iter}+1$.
\UNTIL
$\mathrm{iter} = \mathrm{iter}_\mathrm{max}$ or ${\left\| {\mathbf{{p}}^{\mathrm{iter}} - \mathbf{{p}}^{\mathrm{iter}-1}} \right\|}\le \epsilon$.
\STATE Return the solution $\mathbf{{p}}^{*} = \mathbf{{p}}^{\mathrm{iter}}$.
\end{algorithmic}
\end{algorithm}
\vspace{-17mm}
\end{table}

Now, the problem in \eqref{ResourceAllocation_Power2} is a convex programming problem which can be solved efficiently by standard convex problem solvers, such as CVX \cite{cvx}.
Based on D.C. programming \cite{VucicProofDC}, an iteration algorithm is developed to tighten the obtained upper bound in \eqref{ResourceAllocation_Power2}, which is shown in \textbf{Algorithm} \ref{alg2}.
The power allocation algorithm is initialized with ${\bf{p}}^{1}$, which is obtained by solving the problem in \eqref{ResourceAllocation_Power2} with ${H_1}\left( {\bf{p}} \right)$ as the objective function.
In the $\mathrm{iter}$-th iteration, the updated solution $\mathbf{{p}}^{\mathrm{iter}+1}$ is obtained by solving the problem in \eqref{ResourceAllocation_Power2} with $\mathbf{{p}}^{\mathrm{iter}}$.
The algorithm will terminate when the maximum iteration number is reached, i.e., $\mathrm{iter} = \mathrm{iter}_\mathrm{max}$, or the change of power allocation solutions between adjacent iterations becomes smaller than a given convergence tolerance, i.e., ${\left\| {\mathbf{{p}}^{\mathrm{iter}} - \mathbf{{p}}^{\mathrm{iter}-1}} \right\|}\le \epsilon$.
Note that, with differentiable convex functions ${H_1}\left( {\bf{p}} \right)$ and ${H_2}\left( {\bf{p}} \right)$, the proposed power allocation algorithm converges to a stationary point with a polynomial time computational complexity\cite{VucicProofDC}.

\vspace*{-2mm}
\section{Simulation Results}

In this section, we evaluate the performance of our proposed multi-beam mmWave-NOMA scheme via simulations.
Unless specified otherwise, the simulation setting is given as follows.
We consider an mmWave system with carrier frequency at $28$ GHz.
There are one LOS path and $L = 10$ NLOS paths for the channel model in \eqref{ChannelModel1} and the path loss models for LOS and NLOS paths follow Table I in \cite{AkdenizChannelMmWave}.
A single hexagonal cell with a BS located at the cell center with a cell radius of $200$ m is considered.
All the $K$ users are randomly and uniformly distributed in the cell unless further specified.
The maximum transmit power of the BS is 46 dBm, i.e., $p_{\mathrm{BS}} \le 46$ dBm and the noise power at all the users is assumed identical with $\sigma^2 = -80$ dBm.
We assume that there are $M_{\mathrm{BS}} = 100$ antennas equipped at the BS and $M_{\mathrm{UE}} = 10$ antennas equipped at each user terminals.
The minimum number of antennas allocated to each user is assumed as 10\% of $M_{\mathrm{BS}}$, i.e., ${M_{\min }} = \frac{1}{10} M_{\mathrm{BS}}$.
The minimum rate requirement ${R_{\min }}$ is selected from a uniform distributed random variable with the range of $(0,5]$ bit/s/Hz.
The simulation results shown in the sequel are averaged over different realizations of the large scaling fading, the small scale fading, and the minimum rate data requirement.

To show the effectiveness of our proposed two-stage resource allocation design, we compare the performance in the two stages to their corresponding optimal benchmark, respectively.
In particular, we compare the performance of our proposed coalition formation algorithm to the optimal exhaustive user grouping and antenna allocation.
Given the same obtained user grouping and antenna allocation strategy in the first stage, the performance of our proposed digital precoder and power allocation is compared to the optimal dirty paper coding (DPC) scheme\cite{VishwanathDuality} in the second stage.
On the other hand, to demonstrate the advantages of our proposed multi-beam mmWave-NOMA scheme, we consider two baseline schemes in our simulations.
For baseline 1, the conventional mmWave-OMA scheme is considered where each RF chain can be allocated to at most one user.
To accommodate all the $K$ users on $N_{\mathrm{RF}}$ RF chains with $N_{\mathrm{RF}} \le K \le 2N_{\mathrm{RF}}$, users are scheduled into two time slots.
In each time slot, we perform the downlink mmWave-OMA transmission via a ZF digital precoder and a power allocation design without intra-group interference terms and constraint C3 in \eqref{ResourceAllocation_Power}.
Note that, for a fair comparison, a user can be allocated to at most one time slot in our considered mmWave-OMA scheme since a user can only be associated with at most one RF chain in our proposed multi-beam mmWave-NOMA scheme.
For baseline 2, the single-beam mmWave-NOMA scheme is considered where only the users' LOS AOD within the same $-3$ dB main beamwidth can be considered as a NOMA group\footnote{It has been demonstrated that, in mmWave systems, the angle difference based user pairing \cite{zhouperformance} is superior to the channel gain based user pairing as in conventional NOMA schemes\cite{Dingtobepublished}.}.
If the resulting number of users and single-beam NOMA groups is equal or smaller than $N_{\mathrm{RF}}$, mmWave-OMA transmission is used for spatial multiplexing.
Otherwise, all the users and single-beam NOMA groups are scheduled into two time slots and then we perform mmWave-OMA transmission on each time slot.
For a fair comparison, both our proposed scheme and the baseline schemes are based on the LOS CSI only, i.e., $\left\{ {\theta _{1,0}}, \ldots ,{\theta _{K,0}} \right\}$ and $\left\{ {\alpha _{1,0}}, \ldots ,{\alpha _{K,0}} \right\}$.

\subsection{Convergence of the Proposed Coalition Formation Algorithm}
\begin{figure}[t!]
\centering\vspace{-7mm}
\includegraphics[width=3.5in]{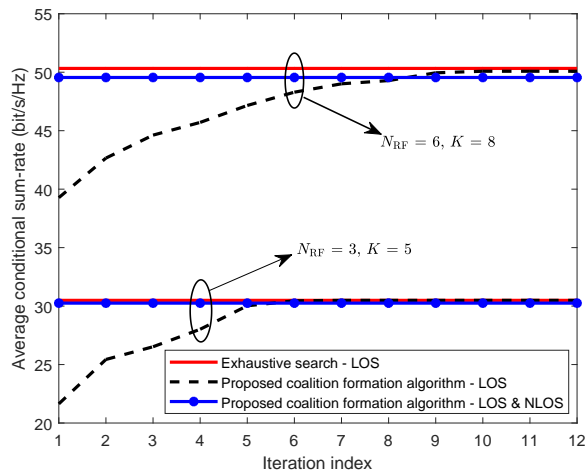}\vspace{-5mm}
\caption{Convergence of our proposed coalition formation algorithm in the first stage.}\vspace{-9mm}
\label{Convergence}
\end{figure}

Fig. \ref{Convergence} illustrates the average conditional system sum-rate in the first stage versus the iteration index to show the convergence of our proposed coalition formation algorithm for user grouping and antenna allocation.
The performance of the exhaustive search on user grouping and antenna allocation is also shown as a benchmark.
Due to the large computational complexity of the optimal exhaustive search, we consider two simulation cases with ${N_{{\mathrm{RF}}}} = 3, K = 5$ and ${N_{{\mathrm{RF}}}} = 6, K = 8$.
The BS transmit power is set as $p_{\mathrm{BS}} = 30$ dBm.
Note that our proposed coalition formation algorithm is applicable to the case with a larger number of RF chains and users as shown in the following simulation cases.
We can observe that the average conditional system sum-rate of our proposed coalition formation algorithm monotonically increases with the iteration index.
Particularly, it can converge to a close-to-optimal performance compared to the exhaustive search within only $10$ iterations on average.
This demonstrates the fast convergence and the effectiveness of our proposed coalition formation algorithm.
With the user grouping and antenna allocation strategy obtained from our proposed coalition formation algorithm, the average conditional sum-rate performance of a multipath channel with both LOS and NLOS paths is also shown.
It can be observed that the performance degradation due to the ignorance of NLOS paths information in the first stage is very limited, especially for the case with small numbers of RF chains and users.
In fact, the channel gain of the LOS path is usually much stronger than that of the NLOS paths in mmWave frequency bands due to the high attenuation in reflection and penetration, c.f. \cite{Rappaport2013,WangBeamSpace2017}.
Besides, the analog beamforming of the massive antennas array at the BS can focus the energy on the LOS AOD and reduce the signal leakage to the NLOS AODs, and hence further reduce the impact of NLOS paths on the system performance.

\subsection{Average System Sum-rate versus Transmit Power at the BS}
\begin{figure}[t!]
\centering\vspace{-7mm}
\includegraphics[width=3.5in]{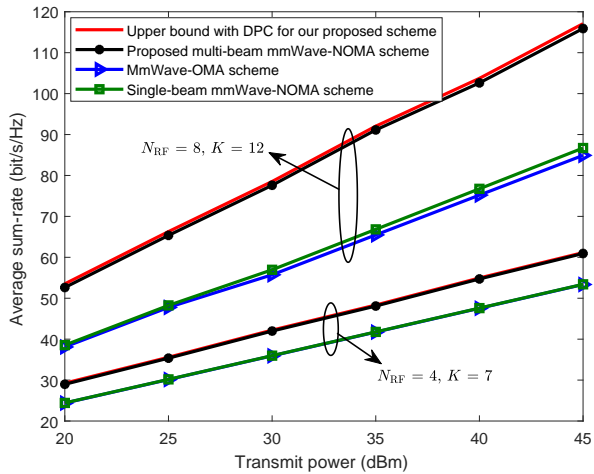}\vspace{-5mm}
\caption{Average system sum-rate (bit/s/Hz) versus the transmit power (dBm) at the BS.}\vspace{-9mm}
\label{Case2}
\end{figure}

Fig. \ref{Case2} illustrates the average system sum-rate versus the total transmit power  $p_{\mathrm{BS}}$ at the BS.
The performance for our proposed scheme with the optimal DPC in the second stage is also shown as the performance benchmark.
Two baseline schemes are considered for comparison and two simulation cases with ${N_{{\mathrm{RF}}}} = 4, K = 7$ and ${N_{{\mathrm{RF}}}} = 8, K = 12$ are included.
We observe that the average system sum-rate monotonically increases with the transmit power since the proposed algorithm can efficiently allocate the transmit power when there is a larger transmit power budget.
Besides, it can be observed that the performance of our proposed resource allocation scheme can approach its upper bound achieved by DPC in the second stage.
It is owing to the interference management capability of our proposed resource allocation design.
In particular, the designed user grouping and antenna allocation algorithm is able to exploit the users' AOD distribution to avoid a large inter-group interference.
Besides, the adopted ZF digital precoder can further suppress the inter-group interference.
Within each NOMA group, the intra-group interference experienced at the strong user can be controlled with the SIC decoding and the intra-group interference at the weak user is very limited owing to our proposed power allocation design.

Compared to the existing single-beam mmWave-NOMA scheme and the mmWave-OMA scheme, the proposed multi-beam mmWave-NOMA scheme can provide a higher spectral efficiency.
This is because that the proposed multi-beam mmWave-NOMA scheme is able to pair two NOMA users with arbitrary AODs, which can generate more NOMA groups and exploit the multi-user diversity more efficiently.
We note that the performance of the single-beam mmWave-NOMA scheme is only slightly better than that of the mmWave-OMA scheme.
It is due to the fact that the probability of multiple users located in the same analog beam is very low\cite{zhouperformance}.
Note that, the average system sum-rate of mmWave systems is much larger than the typical value in microwave systems\cite{WeiTCOM2017}.
It is due to the array gain brought by the large number of antennas equipped at both the BS and user terminals\footnote{We note that the average sum-rate per user obtained in our simulations is comparable to the simulation results in the literature in the field of mmWave communications \cite{AlkhateebPrecoder2015,zhao2017multiuser}.}.

\subsection{Average System Sum-rate versus Number of Antennas at the BS}
\begin{figure}[t!]
\centering\vspace{-7mm}
\includegraphics[width=3.5in]{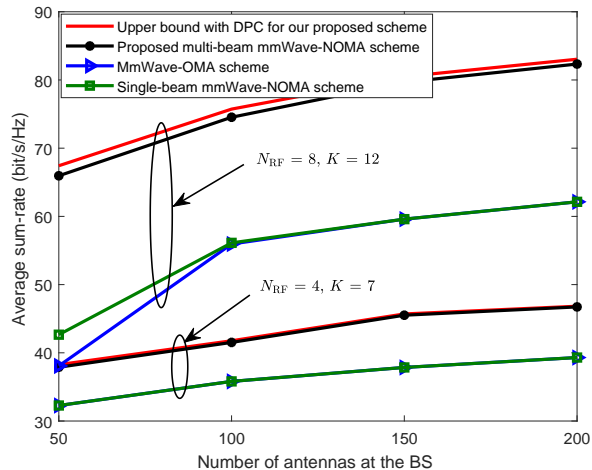}\vspace{-5mm}
\caption{Average system sum-rate (bit/s/Hz) versus the number of antennas equipped at the BS.}\vspace{-9mm}
\label{Case3}
\end{figure}

Fig. \ref{Case3} illustrates the average system sum-rate versus the number of antennas  $M_{\mathrm{BS}}$ equipped at the BS.
Note that, we fix the number of RF chains and only vary the number of antennas equipped at the BS ranging from $50$ to $200$ for the considered hybrid mmWave system.
The simulation setup is the same as Fig. \ref{Case2}, except that we fix the transmit power as $p_{\mathrm{BS}} = 30$ dBm.
We observe that the average system sum-rate increases monotonically with the number of antennas equipped at the BS due to the increased array gain.
Compared to the two baseline schemes, a higher spectral efficiency can be achieved by the proposed multi-beam NOMA scheme due to its higher flexibility of user pairing and enabling a more efficient exploitation of multi-user diversity.
In addition, it can be observed that the performance of the single-beam mmWave-NOMA scheme is almost the same as that of the mmWave-OMA scheme and it is only slightly better than that of the mmWave-OMA scheme when there is a small number of antennas.
In fact, the beamwidth is larger for a small number of antennas, which results in a higher probability for serving multiple NOMA users via the same analog beam.
Therefore, only in the case of a small number of antennas, the single-beam mmWave-NOMA scheme can form more NOMA groups and can provide a higher spectral efficiency than the mmWave-OMA scheme.

\subsection{Average System Sum-rate versus User Density}

\begin{figure}[ht]
\begin{minipage}{.47\textwidth}
\centering\vspace{-6mm}
\includegraphics[width=2.1in]{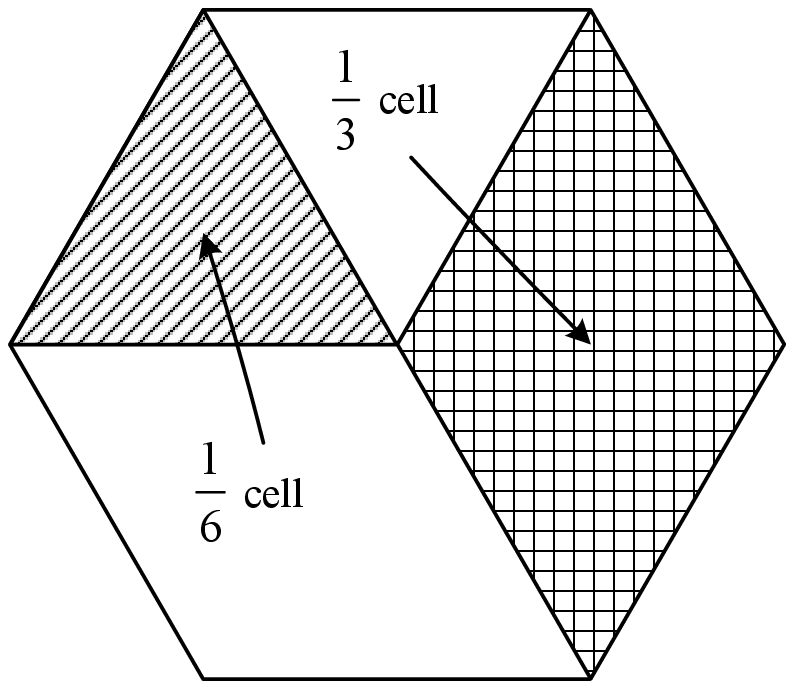}
\caption{An illustration of the selected portions of a cell.}\vspace{-9mm}
\label{SelectedPortions}
\end{minipage}
\hspace*{1.5mm}
\begin{minipage}{.47\textwidth}
\centering\vspace{-6mm}
\includegraphics[width=3.0in]{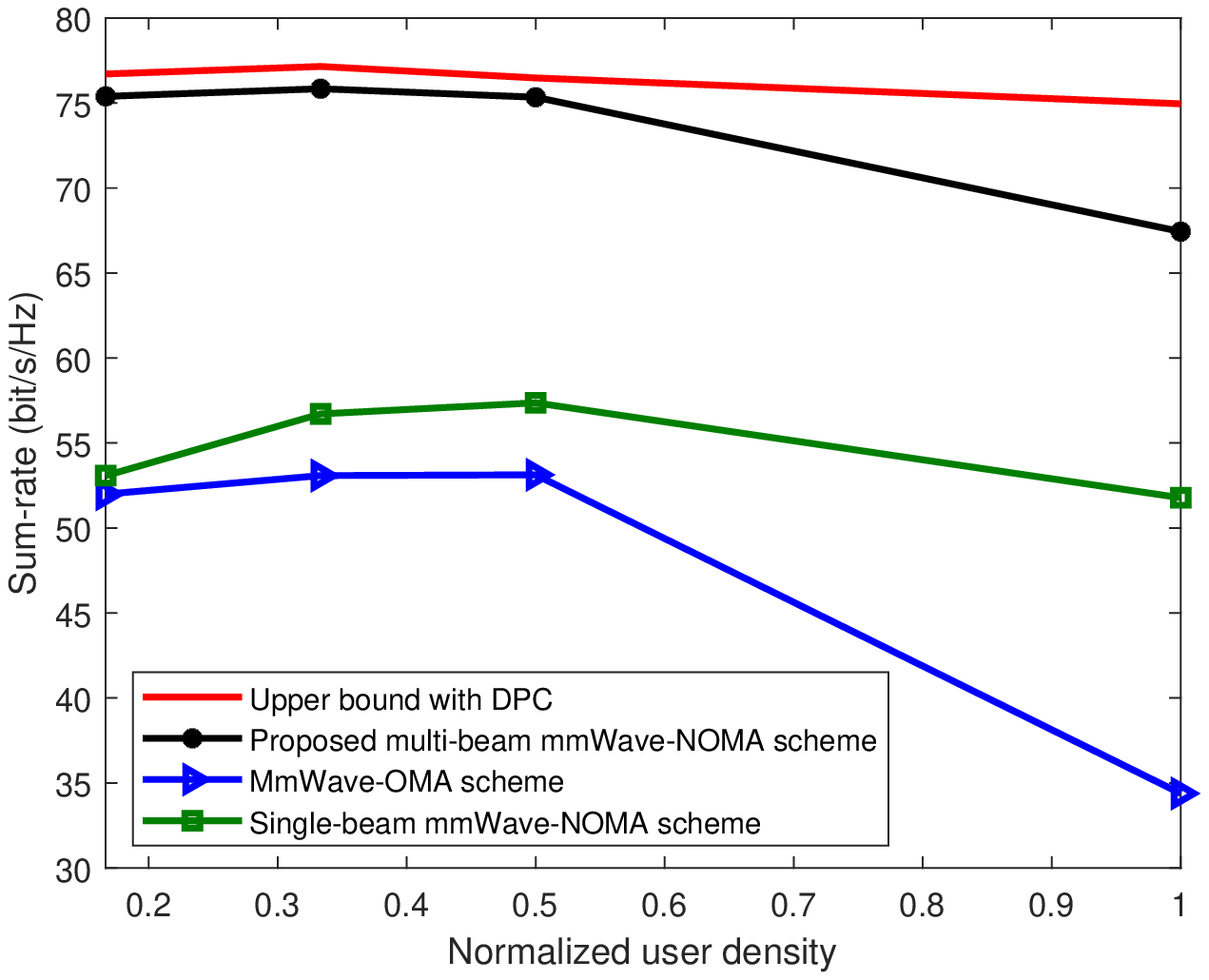}\vspace{-5mm}
\caption{Average system sum-rate (bit/s/Hz) versus the normalized user density.}\vspace{-9mm}
\label{Case4}
\end{minipage}
\end{figure}

Recall that the performance of the baseline single-beam mmWave-NOMA scheme is related to all the users' AOD distribution depending on the user density.
It is interesting to investigate the relationship between the system performance and the user density $\rho = \frac{K}{\mathrm{Area}}$, where $\mathrm{Area}$ denotes the area of the selected portions in the cell.
Fig. \ref{Case4} illustrates the average system sum-rate versus the normalized user density.
To facilitate the simulations, we keep the number of RF chains as ${N_{{\mathrm{RF}}}} = 8$ and the number of users as $K = 12$ and change the area by selecting different portions of the cell, as illustrated in Fig. \ref{SelectedPortions}.
In the selected area, all the $K$ users are uniformly and randomly deployed.
For instance, the minimum normalized user density is obtained with all the $K$ users randomly scattered in the whole hexagonal cell, while the maximum normalized user density is obtained with all the $K$ users randomly deployed in the $\frac{1}{6}$ cell.
The total transmit power at the BS is $p_{\mathrm{BS}} = 30$ dBm.
We can observe that the average system sum-rate of the mmWave-OMA scheme decreases with the user density due to the high channel correlation among users, which introduces a higher inter-user interference.
In addition, the average system sum-rate of the single-beam mmWave-NOMA scheme firstly increases and then decreases with user density.
It is because that, in the low density regime, increasing the user density can provide a higher probability of multiple users located in the same analog beam and thus more NOMA groups can be formed.
On the other hand, in the high density regime, the inter-group interference becomes more severe with the increasing user density.
Besides, it can be observed that our proposed multi-beam mmWave-NOMA scheme can offer a higher average system sum-rate compared to the baseline schemes in the whole user density range.
Note that although the performance of the proposed scheme decreases with the user density, it decreases much slower than that of the mmWave-OMA scheme.
In fact, the inter-group interference becomes severe when more than two users are located in the same analog beam due to the high user density.
However, the proposed scheme can still exploit the multi-user diversity and provide a substantial system performance gain compared to the mmWave-OMA scheme.

\section{Conclusion}
In this paper, we proposed a multi-beam NOMA framework for hybrid mmWave systems and studied the resource allocation design for the proposed multi-beam mmWave-NOMA scheme.
In particular, a beam splitting technique was proposed to generate multiple analog beams to serve multiple users for NOMA transmission.
Our proposed multi-beam mmWave-NOMA scheme is more practical than the conventional single-beam mmWave-NOMA schemes, which can flexibly pair NOMA users with an arbitrary AOD distribution.
As a result, the proposed scheme can generate more NOMA groups and hence is able to explore the multi-user diversity more efficiently.
%
To unlock the potential of the proposed multi-beam NOMA scheme, a two-stage resource allocation was designed to improve the system performance.
More specific, a coalition formation algorithm based on coalition formation game theory was developed for user grouping and antenna allocation in the first stage, while a power allocation based on a ZF digital precoder was proposed to maximize the system sum-rate in the second stage.
Simulation results demonstrated that the proposed resource allocation can achieve a close-to-optimal performance in each stage and our proposed multi-beam mmWave NOMA scheme can offer a substantial system sum-rate improvement compared to the conventional mmWave-OMA scheme and the single-beam mmWave-NOMA scheme.


\begin{thebibliography}{10}
\providecommand{\url}[1]{#1}
\csname url@samestyle\endcsname
\providecommand{\newblock}{\relax}
\providecommand{\bibinfo}[2]{#2}
\providecommand{\BIBentrySTDinterwordspacing}{\spaceskip=0pt\relax}
\providecommand{\BIBentryALTinterwordstretchfactor}{4}
\providecommand{\BIBentryALTinterwordspacing}{\spaceskip=\fontdimen2\font plus
\BIBentryALTinterwordstretchfactor\fontdimen3\font minus
  \fontdimen4\font\relax}
\providecommand{\BIBforeignlanguage}[2]{{%
\expandafter\ifx\csname l@#1\endcsname\relax
\typeout{** WARNING: IEEEtran.bst: No hyphenation pattern has been}%
\typeout{** loaded for the language `#1'. Using the pattern for}%
\typeout{** the default language instead.}%
\else
\language=\csname l@#1\endcsname
\fi
#2}}
\providecommand{\BIBdecl}{\relax}
\BIBdecl

\bibitem{Wei2018mmWaveNOMA}
Z.~Wei, L.~Zhao, J.~Guo, D.~W.~K. Ng, and J.~Yuan, ``A multi-beam {NOMA}
  framework for hybrid mmwave systems,'' \emph{arXiv preprint
  arXiv:1804.08303}, accepted, ICC 2018.

\bibitem{wong2017key}
V.~W. Wong, R.~Schober, D.~W.~K. Ng, and L.-C. Wang, \emph{Key Technologies for
  {5G} Wireless Systems}.\hskip 1em plus 0.5em minus 0.4em\relax Cambridge
  University Press, 2017.

\bibitem{Rappaport2013}
T.~Rappaport, S.~Sun, R.~Mayzus, H.~Zhao, Y.~Azar, K.~Wang, G.~Wong, J.~Schulz,
  M.~Samimi, and F.~Gutierrez, ``Millimeter wave mobile communications for {5G}
  cellular: {It} will work!'' \emph{IEEE Access}, vol.~1, pp. 335--349, May
  2013.

\bibitem{Dai2015}
L.~Dai, B.~Wang, Y.~Yuan, S.~Han, I.~Chih-Lin, and Z.~Wang, ``Non-orthogonal
  multiple access for {5G}: solutions, challenges, opportunities, and future
  research trends,'' \emph{IEEE Commun. Mag.}, vol.~53, no.~9, pp. 74--81, Sep.
  2015.

\bibitem{Ding2015b}
Z.~Ding, Y.~Liu, J.~Choi, Q.~Sun, M.~Elkashlan, C.~L. I, and H.~V. Poor,
  ``Application of non-orthogonal multiple access in {LTE} and {5G} networks,''
  \emph{IEEE Commun. Mag.}, vol.~55, no.~2, pp. 185--191, Feb. 2017.

\bibitem{WeiSurvey2016}
Z.~Wei, Y.~Jinhong, D.~W.~K. Ng, M.~Elkashlan, and Z.~Ding, ``A survey of
  downlink non-orthogonal multiple access for {5G} wireless communication
  networks,'' \emph{ZTE Commun.}, vol.~14, no.~4, pp. 17--25, Oct. 2016.

\bibitem{XiaoMing2017}
M.~Xiao, S.~Mumtaz, Y.~Huang, L.~Dai, Y.~Li, M.~Matthaiou, G.~K. Karagiannidis,
  E.~Bj{\"{o}}rnson, K.~Yang, C.~L. I, and A.~Ghosh, ``Millimeter wave
  communications for future mobile networks,'' \emph{IEEE J. Select. Areas
  Commun.}, vol.~35, no.~9, pp. 1909--1935, Sep. 2017.

\bibitem{zhao2017multiuser}
L.~Zhao, D.~W.~K. Ng, and J.~Yuan, ``Multi-user precoding and channel
  estimation for hybrid millimeter wave systems,'' \emph{IEEE J. Select. Areas
  Commun.}, vol.~35, no.~7, Jul. 2017.

\bibitem{GaoSubarray}
X.~Gao, L.~Dai, S.~Han, C.~L. I, and R.~W. Heath, ``Energy-efficient hybrid
  analog and digital precoding for {MmWave} {MIMO} systems with large antenna
  arrays,'' \emph{IEEE J. Select. Areas Commun.}, vol.~34, no.~4, pp.
  998--1009, Apr. 2016.

\bibitem{lin2016energy}
C.~Lin and G.~Y. Li, ``Energy-efficient design of indoor mmwave and {sub-THz}
  systems with antenna arrays,'' \emph{IEEE Trans. Wireless Commun.}, vol.~15,
  no.~7, pp. 4660--4672, Mar. 2016.

\bibitem{van2002optimum}
H.~L. Van~Trees, \emph{Optimum array processing: Part {IV} of detection,
  estimation and modulation theory}.\hskip 1em plus 0.5em minus 0.4em\relax
  Wiley Online Library, 2002, vol.~1.

\bibitem{zhang2016energy}
Y.~Zhang, H.~M. Wang, T.~X. Zheng, and Q.~Yang, ``Energy-efficient transmission
  design in non-orthogonal multiple access,'' \emph{IEEE Trans. Veh. Technol.},
  vol.~66, no.~3, pp. 2852--2857, Mar. 2016.

\bibitem{Andrews2014}
J.~Andrews, S.~Buzzi, W.~Choi, S.~Hanly, A.~Lozano, A.~Soong, and J.~Zhang,
  ``What will {5G} be?'' \emph{IEEE J. Select. Areas Commun.}, vol.~32, no.~6,
  pp. 1065--1082, Jun. 2014.

\bibitem{Ding2014}
Z.~Ding, Z.~Yang, P.~Fan, and H.~Poor, ``On the performance of non-orthogonal
  multiple access in {5G} systems with randomly deployed users,'' \emph{IEEE
  Signal Process. Lett.}, vol.~21, no.~12, pp. 1501--1505, Dec. 2014.

\bibitem{He2017}
B.~He, A.~Liu, N.~Yang, and V.~K.~N. Lau, ``On the design of secure
  non-orthogonal multiple access systems,'' \emph{IEEE J. Select. Areas
  Commun.}, vol.~35, no.~10, pp. 2196--2206, Oct. 2017.

\bibitem{Ding2017RandomBeamforming}
Z.~Ding, P.~Fan, and H.~V. Poor, ``Random beamforming in millimeter-wave {NOMA}
  networks,'' \emph{IEEE Access}, vol.~5, pp. 7667--7681, Feb. 2017.

\bibitem{Cui2017Optimal}
J.~Cui, Y.~Liu, Z.~Ding, P.~Fan, and A.~Nallanathan, ``Optimal user scheduling
  and power allocation for millimeter wave {NOMA} systems,'' \emph{IEEE Trans.
  Wireless Commun.}, vol.~17, no.~3, pp. 1502--1517, Mar. 2018.

\bibitem{WangBeamSpace2017}
B.~Wang, L.~Dai, Z.~Wang, N.~Ge, and S.~Zhou, ``Spectrum and energy efficient
  beamspace {MIMO-NOMA} for millimeter-wave communications using lens antenna
  array,'' \emph{IEEE J. Select. Areas Commun.}, vol.~PP, no.~99, pp. 1--1,
  Jul. 2017.

\bibitem{Xiao2017MultiBeam}
Z.~Xiao, L.~Dai, Z.~Ding, J.~Choi, and P.~Xia, ``Millimeter-wave communication
  with non-orthogonal multiple access for {5G},'' \emph{arXiv preprint
  arXiv:1709.07980}, 2017.

\bibitem{Zhu2017Joint}
Z.~Xiao, L.~Zhu, J.~Choi, P.~Xia, and X.~G. Xia, ``Joint power allocation and
  beamforming for non-orthogonal multiple access {(NOMA)} in {5G}
  millimeter-wave communications,'' \emph{IEEE Trans. Wireless Commun.}, 2018,
  early access.

\bibitem{SaadCoalitionalGame}
W.~Saad, Z.~Han, M.~Debbah, A.~Hjorungnes, and T.~Basar, ``Coalitional game
  theory for communication networks,'' \emph{IEEE Signal Process. Mag.},
  vol.~26, no.~5, pp. 77--97, Sep. 2009.

\bibitem{SaadCoalitional2012}
W.~Saad, Z.~Han, R.~Zheng, A.~Hjorungnes, T.~Basar, and H.~V. Poor,
  ``Coalitional games in partition form for joint spectrum sensing and access
  in cognitive radio networks,'' \emph{IEEE J. Select. Topics Signal Process.},
  vol.~6, no.~2, pp. 195--209, Apr. 2012.

\bibitem{WangCoalitionNOMA}
K.~Wang, Z.~Ding, and W.~Liang, ``A game theory approach for user grouping in
  hybrid non-orthogonal multiple access systems,'' in \emph{Proc. IEEE Intern.
  Sympos. on Wireless Commun. Systems}, Sep. 2016, pp. 643--647.

\bibitem{SaadCoalitionOrder}
W.~Saad, Z.~Han, M.~Debbah, and A.~Hjorungnes, ``A distributed coalition
  formation framework for fair user cooperation in wireless networks,''
  \emph{IEEE Trans. Wireless Commun.}, vol.~8, no.~9, pp. 4580--4593, Sep.
  2009.

\bibitem{Han2012}
Z.~Han, D.~Niyato, W.~Saad, T.~Baar, and A.~Hjrungnes, \emph{Game Theory in
  Wireless and Communication Networks: Theory, Models, and Applications}.\hskip
  1em plus 0.5em minus 0.4em\relax New York, NY, USA: Cambridge University
  Press, 2012.

\bibitem{BeamTracking}
V.~Va, H.~Vikalo, and R.~W. Heath, ``Beam tracking for mobile millimeter wave
  communication systems,'' in \emph{Proc. IEEE Global Conf. on Signal and Inf.
  Process.}, Dec. 2016, pp. 743--747.

\bibitem{WeiTCOM2017}
Z.~Wei, D.~W.~K. Ng, J.~Yuan, and H.~M. Wang, ``Optimal resource allocation for
  power-efficient {MC-NOMA} with imperfect channel state information,''
  \emph{IEEE Trans. Commun.}, vol.~PP, no.~99, pp. 1--1, May 2017.

\bibitem{Sun2016Fullduplex}
Y.~Sun, D.~W.~K. Ng, Z.~Ding, and R.~Schober, ``Optimal joint power and
  subcarrier allocation for full-duplex multicarrier non-orthogonal multiple
  access systems,'' \emph{IEEE Trans. Commun.}, vol.~65, no.~3, pp. 1077--1091,
  Mar. 2017.

\bibitem{Sohrabi2016}
F.~Sohrabi and W.~Yu, ``Hybrid digital and analog beamforming design for
  large-scale antenna arrays,'' \emph{IEEE J. Select. Areas Commun.}, vol.~10,
  no.~3, pp. 501--513, Apr. 2016.

\bibitem{AlkhateebPrecoder2015}
A.~Alkhateeb, G.~Leus, and R.~W. Heath, ``Limited feedback hybrid precoding for
  multi-user millimeter wave systems,'' \emph{IEEE Trans. Wireless Commun.},
  vol.~14, no.~11, pp. 6481--6494, Nov. 2015.

\bibitem{Hanif2016}
M.~F. Hanif, Z.~Ding, T.~Ratnarajah, and G.~K. Karagiannidis, ``A
  minorization-maximization method for optimizing sum rate in the downlink of
  non-orthogonal multiple access systems,'' \emph{IEEE Trans. Signal Process.},
  vol.~64, no.~1, pp. 76--88, Jan. 2016.

\bibitem{Sun2017MIMONOMA}
Y.~Sun, D.~W.~K. Ng, and R.~Schober, ``Optimal resource allocation for
  multicarrier {MISO-NOMA} systems,'' in \emph{Proc. IEEE Intern. Commun.
  Conf.}, May 2017, pp. 1--7.

\bibitem{Mumtaz2016mmwave}
S.~Mumtaz, J.~Rodriguez, and L.~Dai, \emph{MmWave Massive {MIMO}: {A} Paradigm
  for {5G}}.\hskip 1em plus 0.5em minus 0.4em\relax Academic Press, 2016.

\bibitem{Boyd2004}
S.~Boyd and L.~Vandenberghe, \emph{Convex optimization}.\hskip 1em plus 0.5em
  minus 0.4em\relax Cambridge university press, 2004.

\bibitem{cvx}
M.~Grant and S.~Boyd, ``{CVX}: Matlab software for disciplined convex
  programming, version 2.1,'' \url{http://cvxr.com/cvx}, Mar. 2014.

\bibitem{VucicProofDC}
N.~Vucic, S.~Shi, and M.~Schubert, ``{DC} programming approach for resource
  allocation in wireless networks,'' in \emph{Proc. Int. Symp. Model. Optim.
  Mobile Ad Hoc Wireless Netw.}, May 2010, pp. 380--386.

\bibitem{AkdenizChannelMmWave}
M.~R. Akdeniz, Y.~Liu, M.~K. Samimi, S.~Sun, S.~Rangan, T.~S. Rappaport, and
  E.~Erkip, ``Millimeter wave channel modeling and cellular capacity
  evaluation,'' \emph{IEEE J. Select. Areas Commun.}, vol.~32, no.~6, pp.
  1164--1179, Jun. 2014.

\bibitem{VishwanathDuality}
S.~Vishwanath, N.~Jindal, and A.~Goldsmith, ``Duality, achievable rates, and
  sum-rate capacity of gaussian {MIMO} broadcast channels,'' \emph{IEEE Trans.
  Inf. Theory}, vol.~49, no.~10, pp. 2658--2668, Oct. 2003.

\bibitem{zhouperformance}
Y.~Zhou, V.~W. Wong, and R.~Schober, ``Performance analysis of millimeter wave
  {NOMA} networks with beam misalignment,'' accepted, ICC 2018.

\bibitem{Dingtobepublished}
Z.~Ding, P.~Fan, and H.~V. Poor, ``Impact of user pairing on {5G} nonorthogonal
  multiple-access downlink transmissions,'' \emph{IEEE Trans. Veh. Technol.},
  vol.~65, no.~8, pp. 6010--6023, Aug. 2016.

\end{thebibliography}


\end{document}